# Astronomical alignments as the cause of ~M6+ seismicity

Mensur Omerbashich
*The Bosnian Royal Family, P.O. Box 1, Sarajevo, Bosnia*
*The European Royal Society, omerbashich@theroyalsociety.eu; cc: omerbashich@gmail.com*

I here demonstrate empirically my georesonator concept in which tidally induced magnification of Earth masses' resonance causes seismicity on Earth. To that end, I show that all strong (~M6+) earthquakes of 2010 occurred during Earth's long ($\Delta t$ > 3 days) astronomical alignments within our solar system. I then show that the same holds true for all very strong (~M8+) earthquakes of the decade of 2000's. Finally, the strongest (M8.6+) earthquakes of the past century are shown to have occurred during Earth's multiple long alignments, too, whereas half of the high-strongest (M9+) earthquakes occurred during Full Moon. To add to my solution's robustness, I include alignments to the comet C/2010 X1 (Elenin) as it is the only heavenly body currently in our solar system besides planets, and show that it impacted very strong seismicity since 2007 (and strongest seismicity, perhaps since 1965). I conclude my empirical demonstration with a six-month, 100%-successful forecast of M6+ magnification, performed online in near real-time from October 2011 until April 2012, and which shows always the same, fractal pattern of intensifying-peaking-diminishing of strong earthquakes' strength during long alignments, and the absence of the pattern outside alignments. Days without killer quake can now be known exactly, amounting to several months a year of exact seismic forecast (telling with certainty there will be no strong seisms on that-and-that date). Approximate forecast of strength and timing of strong earthquakes is feasible for the rest of the year, as well as of location once the alignment starts. I thus verified my hyperresonator concept too, arrived at earlier as a mathematical-physical solution to the most general extension of georesonator. As gravity can now be expressed in terms of speed of light, hyperresonance is akin to tuning forks triggering each other at a distance, which infers the existence of aether as required for propagation of mutually tuning (or, interfering) oscillations in Space.

## Introduction

Geophysics cannot explain the mechanism that supplies most of the energy required for tectonogenesis and overall seismicity on Earth [1]. On the other hand, Earth's ~M7+ seismicity can arise as a natural response of our planet to its alignments with other celestial objects, Sun/Moon in particular, so that seismicity exhibits up to 3-day phase and maximum displacement up to ~10 m that correspond to M9.5 quakes – the strongest allowed in luni-solar *georesonator* (the Earth thought of as a mechanical oscillator forced by celestial bodies) [2]. That is the extreme case-scenario for our solar system.

I verified the georesonator concept theoretically, by extending it to the most general realm – that of the *hyperresonator* (the Universe thought of as a mechanical oscillator forced by yet outer Universes). As the result, a remarkable mathematical solution was obtained: the first analytical expression for gravitational constant, *G*, was derived, for various scales [3]. Here, instead of regarding just the Moon-Sun forcing, I expand the georesonator so to consider the forcing of the Earth by all celestial objects presently found in our solar system as well.

Units of time are given to within ±1 day, as differential orbital inclinations in the nearly co-planar solar system can be safely ignored. Unless stated otherwise, here by alignments I mean Earth's long alignments ($\Delta t$ > 3 day) with two other bodies in our solar system. The alignments were estimated to ±1 arc°, where spatial angles were expressed in units of time (days). The '~' means ±5%, except for the M6 threshold that was set (when creating the sample) so as to make the sample large enough for depicting a pattern of resonance magnification proposed by [2]. So the M6- earthquakes used in this study were those flagged by USGS as significant for deaths. They are few, so they should not alter the result. Note that the M6 threshold was proper also because the M6+ earthquakes emit ~70%+ of energy released by all earthquakes. This makes the M6.5 a safe choice for threshold in search for an unknown mechanism responsible for most of Earth's tectonogenesis and seismicity. M in this paper marks $M_w$, unless stated otherwise.

## Verification

The georesonator concept is herein verified by comparing the epochs of the last year's strong (~M6+), the last decade's very strong (~M8+), as well as the last century's strongest (M8.6+) seismicity, against the Earth's alignments to significant bodies of mass in our solar system. Here 'significant' are all objects between the Sun and the Neptune's orbit, which can cause gravitational shadowing on the Earth, akin to that caused by the Sun/Moon [2] [3]. So I include the following celestial objects in this verification: the Sun, the Mercury, the Venus, the Mars, the Jupiter, the Saturn, the Uranus, the Neptune, and the comet C/2010X1 (Elenin). The Elenin is included for two main reasons: first, it drags gravitationally locked particles around 30,000+ km across, at the ~1° inclination, making its gravitational shadowing significant for proving the georesonator. Secondly, it passed the Uranus orbit (on the Elenin's own sling-orbit about the Sun) within the previous decade, adding to this verification's robustness in terms of very strong seismicity. From the practical point of view, the inclusion of the Elenin improves robustness of the empirical test while the geophysics community is clueless as to the source of most of the energy supplied to the Earth and required for tectonogenesis (mantle/plates dynamics), as well as for causing strong earthquakes [1]. The georesonator and its generalization, the hyperresonator, account for that universally, i.e., regardless of what heavenly object one is talking about. For example, if we know that volcanoes on Earth erupt with burning-hot lava, but on Titan they spew ice-cold ammonia, the cause of volcanic or, generally, mantle dynamics on Earth cannot lie in thermo-nuclear reactions and processes, but is essentially mechanical.

The use of the previous decade gives ten years of seismic data that are same in kind but entirely independent from the (1990's) data I had used to prove the georesonator mathematically-physically [2] [3]. The JPL Orbit Solutions for C/2010X1 (16 and 23 March 2011) were used for resolving the alignments [4]. The earthquake reference-data are from USGS [5], and lunar phases are from NASA [6]. Then this verification is methodological. No statistical testing was done due to: relatively small sample sizes, diverse distributions [2] and mathematical-physical verification having been achieved already (and via independent reasoning) [3].

Then in order to empirically verify the georesonator concept, it suffices to show that, first, all strong earthquakes of 2010 occurred during alignments and in self-evident fashion (meaning: the kinetic energy always flares in the same fashion, i.e., makes the same kind of serial pattern within a given alignment as well as alignment-to-alignment). Obviously in a forced mechanical oscillator, the pattern takes form of increase-peaking-decrease of energy emissions (here in ~M6+ seismicity). Such a pattern would be the fingerprint of a forced mechanical oscillator at work. Such a pattern would also require for strong quakes to be occurring at remote locations, as the level of energy released lines up in an orderly fashion due to external mechanical forcing of the entire planet i.e. regardless of location. If such a pattern were demonstrated, it would mean that tides do not trigger earthquakes directly (by causing stress/strain release), but that they trigger resonance magnification instead, which then acts as an intermediary giving rise to Earth's shaking (earthquakes), cracking (faults), magma flow (tectonics), and differential core rotation as a downward continuation of the magnified vibration. This was already speculated in [2].

I select year 2010 as it contained the most robust data, supposedly due to the proximity of the comet Elenin. Without a permanent state of unrest, meaning alignments and oscillations going on around the clock, there could be no magnification of such an unrest going on all year round either, which is also why one cannot speak of any correlations/angles whatsoever in this study, let alone search for those. What one can do though is search for the above-described, dominant patterns that are repetitive during a long alignment and from alignment-to-alignment, as the fingerprint of the claimed mechanism. Since there cannot exist any intermediary for transferring the energy between the whole Earth and the Solar system, other than that of the proposed mechanism, if found, the pattern would represent an empirical proof of the mechanism itself. So alignments relevant for seismicity and strong seismicity forecasting are those which are closest in time to the strong earthquake itself, regardless if they occurred "exactly on time" or not, as numerous physical-chemical inner factors affect the timing of magnification and its culmination in particular. That a pattern-making M6+ magnification occurs during a long alignment is far more significant for earthquake forecast than the question of when exactly that alignment begins or ends.

Secondly, one has to show that the very strong decadal earthquakes occurred during long alignments as well. Lastly, the past century's strongest earthquakes must be demonstrated also to have occurred during long alignments as well. Curiously enough, the total number of long alignments and very strong earthquakes was nearly the same over the selected test periods, where 'nearly' depicts the difference ascribable solely to earthquakes of lower magnitude class.

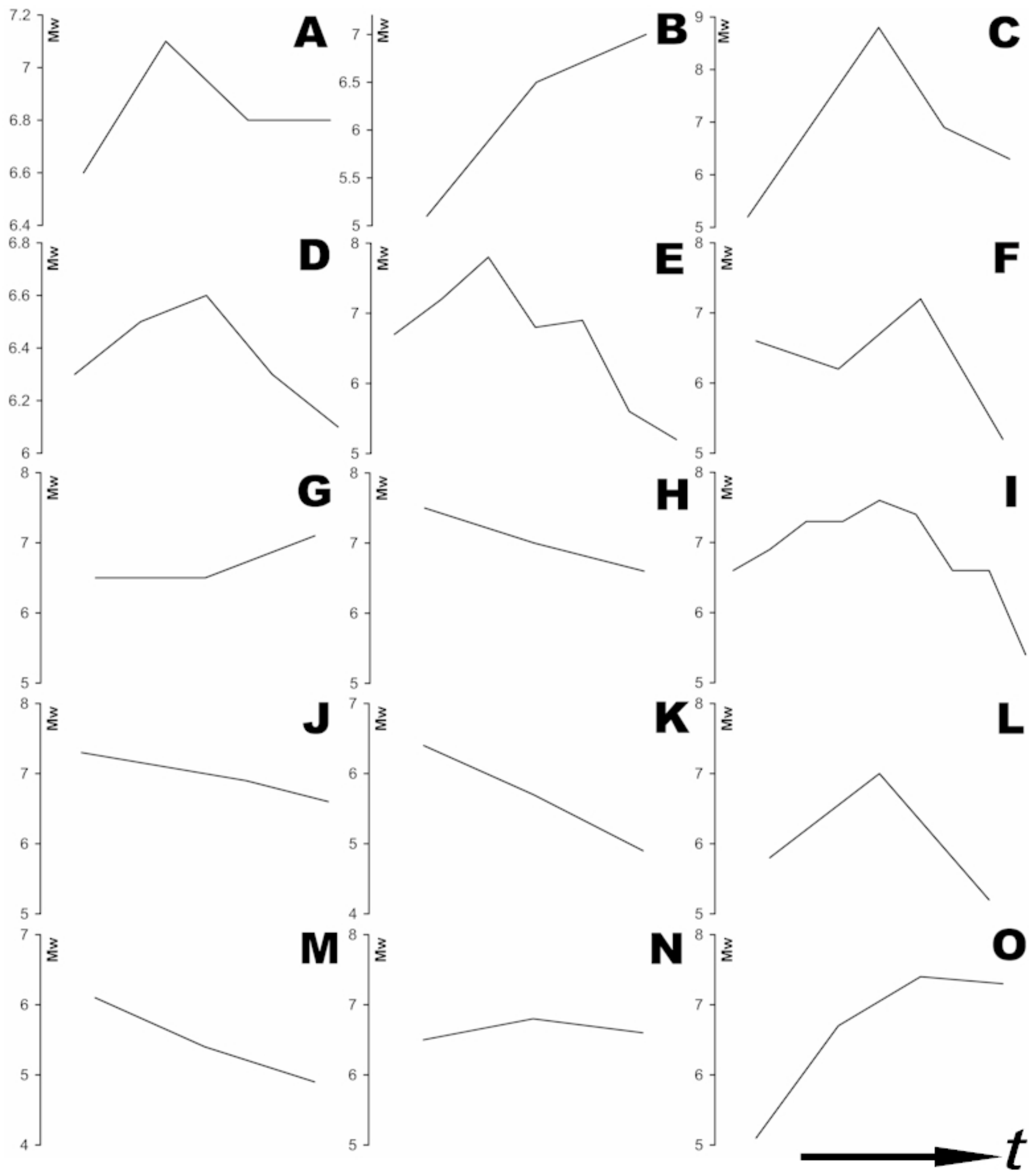

Figure 1. The resonance magnification pattern, as a gravitational shadow traverses the Earth thus disturbing the Earth's masses; cases A-O, Table 1. Resolution increases with the Earth's response as measured by the number of (strong) earthquakes where the total resolvedness is attained past the minimum resolution of 3 (successive strong) earthquakes.

| | ALIGNMENTS | DATE | MOON | LOCATION | MAGNITUDE | DEPTH | LABEL |
|---|---|---|---|---|---|---|---|
| | Earth-Mercury-Sun-Venus | Jan 03 2010 | | Solomon Isles | Mw=6.6 | d=26.0 km | PS |
| | Earth-Mercury-Sun-Venus | Jan 03 2010 | | Solomon Isles | Mw=7.1 | d=25.0 km | PS |
| A | Earth-Mercury-Sun-Venus | Jan 05 2010 | | Sandwich Isles | Mw=6.8 | d=10.0 km | PS |
| | Earth-Mercury-Sun-Venus | Jan 05 2010 | | Solomon Isles | Mw=6.8 | d=15.0 km | PS |
| | Earth-Sun-Venus | Jan 10 2010 | | Indonesia | Mw=5.1 | d=65.0 km | PS |
| B | Earth-Sun-Venus | Jan 10 2010 | | California US | Mw=6.5 | d=29.0 km | PS |
| | Earth-Sun-Venus | Jan 12 2010 | | Haiti | Mw=7.0 | d=13.0 km | PS |
| | Earth proximal to: Mars-Sun-Venus | Jan 17-25 10 | | Yellowstone US | (swarm w/1000s of events) | | PS |
| | Mars-Earth-Venus | Jan 28 2010 | | | | | |
| | Mars-Earth-Sun; **Moon**-Earth-**Sun** | Jan 30 2010 | F | China | Mw=5.1 | d=10.0 km | MPS |
| | Mars-Earth-Mercury; Earth-Moon-**Sun** | Feb 14 2010 | N | | | | |
| | Elenin-Earth-Venus | Feb 18 2010 | | China/RU/N.Korea | Mw=6.9 | d=578 km | CP |
| | Elenin-Earth-Jupiter | Feb 22 2010 | | | | | |
| | Elenin-Earth-Sun | Feb 25 2010 | | China | Mw=5.2 | d=10.0 km | CS |
| | Elenin-Earth-Sun | Feb 26 2010 | | Japan | Mw=7.0 | d=25.0 km | CS |
| C | Elenin-Earth-Sun; Earth-Sun-Jupiter; Sun | Feb 27 2010 | F | Chile | Mw=8.8 | d=23.0 km | CMPS |
| | Elenin-Earth-Sun | Feb 27 2010 | | Chile | Mw=6.9 | d=35.0 km | CS |
| | Elenin-Earth-Sun | Feb 27 2010 | | Argentina | Mw=6.3 | d=10.0 km | CS |
| | Elenin-Earth-Mercury | Mar 04 2010 | | Taiwan | Mw=6.3 | d=21.0 km | CP |
| | Elenin-Earth-Mercury | Mar 04 2010 | | Vanuatu | Mw=6.5 | d=176 km | CP |
| D | | | | | | | |
| | Elenin-Earth-Mercury | Mar 05 2010 | | Chile | Mw=6.6 | d=18.0 km | CP |
| | Elenin-Earth-Mercury | Mar 05 2010 | | Indonesia | Mw=6.3 | d=21.0 km | CP |
| | Elenin-Earth-Mercury; Earth-Sun-Mercury | Mar 08 2010 | | Turkey | Mw=6.1 | d=12.0 km | CPS |
| | Earth-Sun-Mercury; Saturn-Earth-Venus | Mar 11 2010 | | Chile | Mw=6.9 | d=11.0 km | PS |
| | Earth-Sun-Mercury; Saturn-Earth-Venus | Mar 11 2010 | | Chile | Mw=6.7 | d=18.0 km | PS |
| | Earth-Sun-Mercury-Uranus | Mar 14 2010 | | Japan | Mw=6.5 | d=32.0 km | PS |
| | Earth-Sun-Mercury-Uranus; Sun | Mar 16 2010 | N | Chile | Mw=6.7 | d=18.0 km | MPS |
| | Saturn-Earth-Sun | Mar 20 2010 | | Papua NG | Mw=6.6 | d=415 km | PS |
| | Earth-Mercury-Venus; Moon-Earth-Sun | Mar 30 2010 | F | India | Mw=6.7 | d=34.0 km | MPS |
| | Earth-Mercury-Venus | Apr 04 2010 | | Mexico | Mw=7.2 | d= 4.0 km | P |
| | Earth-Mercury-Venus | Apr 06 2010 | | Indonesia | Mw=7.8 | d=31.0 km | P |
| E | Earth-Mercury-Venus | Apr 11 2010 | | Solomon Isles | Mw=6.8 | d=21.0 km | P |
| | Earth-Mercury-Venus; Earth-Moon-Sun | Apr 13 2010 | N | China | Mw=6.9 | d=17.0 km | MPS |
| | Earth-Mercury-Venus | Apr 18 2010 | | Afghanistan | Mw=5.6 | d=13.0 km | P |
| | Earth-Mercury-Venus | Apr 20 2010 | | Australia | Mw=5.2 | d= 0.0 km | P |
| | Saturn-Earth-Uranus | Apr 26 2010 | | Taiwan | Mw=6.5 | d=15.0 km | P |
| | Earth-Mercury-Sun | Apr 30 2010 | | Bering Sea | Mw=6.5 | d=14.0 km | PS |
| | Elenin-Earth-Neptune(Earth-Mercury-Sun) | May 05 2010 | | Indonesia | Mw=6.6 | d=27.0 km | P(S) |
| | Elenin-Earth-Neptune(Earth-Mercury-Sun) | May 06 2010 | | Chile | Mw=6.2 | d=37.0 km | P(S) |
| F | Elenin-Earth-Neptune(Earth-Mercury-Sun) | May 09 2010 | | Indonesia | Mw=7.2 | d=38.0 km | P(S) |
| | Elenin-Earth-Neptune(Earth-Mercury-Sun) | May 14 2010 | N | Algeria | Mw=5.2 | d= 2.0 km | MP(S) |
| | Saturn-Earth-Jupiter | May 24 2010 | | Brazil | Mw=6.5 | d=581 km | P |
| G | Saturn-Earth-Jupiter | May 26 2010 | | Japan | Mw=6.5 | d=10.0 km | P |
| | Saturn-Earth-Jupiter; Moon-Earth-Sun | May 27 2010 | F | Vanuatu | Mw=7.1 | d=31.0 km | MPS |
| | Mars-Earth-Neptune | May 31 2010 | | Andaman, India | Mw=6.5 | d=112 km | P |
| | Earth-Jupiter Uranus; Earth-**Moon**-**Sun** | Jun 12 2010 | N | Nicobar, India | Mw=7.5 | d=35.0 km | MPS |
| H | Earth-Jupiter Uranus | Jun 16 2010 | | Indonesia | Mw=7.0 | d=18.0 km | P |
| | Earth-Jupiter Uranus | Jun 16 2010 | | Indonesia | Mw=6.6 | d=11.0 km | P |
| | Earth-Sun-Mercury; **Moon**-Earth-**Sun** | Jun 26 2010 | F | Solomon Isles | Mw=6.7 | d=35.0 km | MPS |
| | Earth-Sun-Mercury | Jun 30 2010 | | Mexico | Mw=6.3 | d=20.0 km | PS |
| | Earth-Venus-Elenin | Jul 14 2010 | | Chile | Mw=6.6 | d=22.0 km | CP |
| | Saturn-Earth-Uranus | Jul 18 2010 | | Alaska USA | Mw=6.6 | d=14.0 km | P |
| | Saturn-Earth-Uranus | Jul 18 2010 | | Papua NG | Mw=6.9 | d=28.0 km | P |
| | Saturn-Earth-Uranus | Jul 18 2010 | | Papua NG | Mw=7.3 | d=35.0 km | P |
| | Saturn-Earth-Uranus | Jul 23 2010 | | Philippines | Mw=7.3 | d=607 km | P |
| I | Saturn-Earth-Uranus | Jul 23 2010 | | Philippines | Mw=7.6 | d=586 km | P |
| | Saturn-Earth-Uranus | Jul 23 2010 | | Philippines | Mw=7.4 | d=641 km | P |
| | Saturn-Earth-Uranus | Jul 24 2010 | | Philippines | Mw=6.6 | d=553 km | P |
| | Saturn-Earth-Uranus; **Moon**-Earth-**Sun** | Jul 29 2010 | F+3 | Philippines | Mw=6.6 | d=627 km | MPS |
| | Saturn-Earth-Uranus | Jul 30 2010 | | Iran | Mw=5.4 | d=24 km | P |
| | Mars-Earth-Jupiter | Aug 04 2010 | | Papua NG | Mw=6.5 | d=226 km | P |
| | Mars-Earth-Jupiter | Aug 04 2010 | | Papua NG | Mw=7.0 | d=44.0 km | P |
| | Venus-Earth-Jupiter; Earth-**Moon**-**Sun** | Aug 10 2010 | N | Vanuatu | Mw=7.3 | d=25.0 km | MPS |
| J | Venus-Earth-Jupiter | Aug 12 2010 | N+3 | Ecuador | Mw=7.1 | d=207 km | MPS |
| | Venus-Earth-Jupiter | Aug 13 2010 | | Mariana Isles | Mw=6.9 | d=10.0 km | P |
| | Venus-Earth-Jupiter | Aug 14 2010 | | Mariana Isles | Mw=6.6 | d=13.0 km | P |
| | Earth-Venus-Mars Saturn-Earth-Jupiter | Aug 20 2010 | | Papua NG | Mw=6.4 | d=51.0 km | P |
| K | Earth-Venus-Mars | Aug 27 2010 | | Iran | Mw=5.7 | d= 7.0 km | P |
| | Earth-Venus-Mars | Aug 29 2010 | | China | Mw=4.9 | d=53.0 km | P |
| | Earth-Mercury-Sun-Elenin | Sep 03 2010 | | Alaska USA | Mw=6.5 | d=24.0 km | CPS |
| | Earth-Mercury-Sun-Elenin | Sep 03 2010 | | New Zealand | Mw=7.0 | d=12.0 km | CPS |
| | Earth-Venus-Mars | Sep 27 2010 | | Iran | Mw=5.8 | d=27.0 km | P |
| | Earth-Venus-Mars | Sep 29 2010 | | Indonesia | Mw=7.0 | d=26.0 km | P |
| L | Earth-Venus-Mars; Earth-Sun-Saturn; Jupiter-Earth-Mercury | Oct 01 2010 | | | | | |
| | Earth-Venus-Mars | Oct 10 2010 | | Pakistan | Mw=5.2 | d=33.0 km | P |
| | Earth-Sun-Mercury | Oct 21 2010 | | California Gulf | Mw=6.7 | d=10.0 km | PS |
| | Earth-Venus-Mercury | Oct 25 2010 | | Indonesia | Mw=7.8 | d=20.0 km | P |
| | Earth-Venus-Sun | Oct 29 2010 | | | | | |
| | Jupiter-Earth-Elenin | Nov 03 2010 | | Indonesia | Mw=6.1 | d=29.1 km | CP |
| M | Jupiter-Earth-Elenin | Nov 03 2010 | | Serbia | Mw=5.4 | d=10.0 km | CP |
| | Jupiter-Earth-Elenin; Earth-Moon-Sun | Nov 06 2010 | N | Iran | Mw=4.9 | d= 5.0 km | CMP |
| | Elenin-Earth-Uranus | Nov 10 2010 | | India | Mw=6.5 | d=10.0 km | CP |
| | Earth-Mercury-Mars; Moon-Earth-Sun | Nov 21 2010 | F | | | | |
| N | Elenin-Earth-Uranus | Nov 30 2010 | | Japan | Mw=6.8 | d=487 km | CP |
| | Elenin-Earth-Uranus | Dec 02 2010 | | Papua NG | Mw=6.6 | d=33.0 km | CP |
| | Earth-Mercury-Mars | Dec 14 2010 | | | | | |
| | Earth-Mercury-Sun | Dec 19 2010 | | Ethiopia | Mw=5.1 | d=10.0 km | PS |
| O | Earth-Mercury-Sun | Dec 20 2010 | | Iran | Mw=6.7 | d=12.0 km | PS |
| | Earth-Mercury-Sun; **Moon**-Earth-**Sun** | Dec 21 2010 | F | Japan | Mw=7.4 | d=17.0 km | MPS |
| | Earth-Mercury-Sun | Dec 25 2010 | | Vanuatu | Mw=7.3 | d=12.0 km | PS |

Table 1. Comparison of Earth's all long alignments v. all strong (~M6+) earthquakes, of 2010. Earthquake labels: C for Elenin-partaking, M for Moon-partaking, P for planetary-partaking, S for Sun-partaking alignment. Consecutive bold caps on the left hand-side mark the cases of long ($\Delta t$ > 3 days) alignments; see Figure 1. Moon phase labels: 'N' for New Moon, 'F' for Full Moon; both to within ±1 day. Underscored 'Sun' marks a solar eclipse, while underscored 'Moon' marks a lunar eclipse; both of any type and to within 2 lunar months, or within 1 lunar month when in bold.

## Discussion

As seen from Table 1, all strong earthquakes of the year 2010 have occurred during a long alignment. The alignments were occurring at the rate of some 20-30 times per year before the Elenin crossed into the Uranus orbit in 2006, and then around 30-40 times per year afterwards.

Table 1 and Figure 1 indeed do reveal a regular pattern for dissipation of energy in all cases (labeled A-O). The pattern is better resolved for longer lasting alignments, i.e., those that can be related to more than the minimum of three strong earthquakes; see Figure 1. As expected for a georesonator, magnification of Earth masses' oscillation intensifies fastest at the beginning of the alignment, subsequently fading out as the alignment fades out. Then brief alignments, of up to three days in duration, cannot generally be related to strong earthquakes. For example, the killer Haiti earthquake of 12 January came at the culmination of the Earth-Sun-Venus alignment lasting for several days due to the Sun's size.

Focal depth of a strong earthquake is of no relevance, except for some cases of lunar forcing (namely: I, J) [2], indicating that all strong seismicity on Earth is rather due to a general external mechanism, such as the georesonator forced by celestial objects, and not by depth-stratified thermal or material processes. This explains the mysterious occurrence of deep (400+ km) earthquakes.

In order to make the georesonator concept useful, a one-on-one relationship ought to be shown for a minimum of 67% earthquakes potentially dangerous to humans. Therefore, one has to examine how well all of the alignments over a test-period match the very strong seismicity. As required by the georesonator concept, it can be seen from Table 1 that all examined alignments did coincide in time with the occurrence of more than one strong earthquake. Comparison in Table 3, of the JPL alignment data against the USGS annual data averaged over the Centennial Catalog, shows there are on average 20-40 astronomical alignments of any duration, per annum, versus ~150 strong earthquakes per annum. Thus Tables 1-3 essentially establish, at a well over 67% rate, a desired one-on-one relationship between very strong seismicity and long alignments. The matching rate for the testing data was above 90%, which is then adopted hereby as the threshold (declared accuracy) in the approximate earthquake prediction using the georesonator concept. What is the extent and meaning of this approximation?

In the classical view, successful earthquake prediction should provide exact magnitude, time and location of a strong earthquake. However, given that the herein proved georesonator concept actually applies to the physics behind the Earth's strongest seismicity (potentially tragic to humans), the above definition is here regarded impractical and rigid. The definition is then amended to be more humane: –*Successful prediction is one that predicts a very strong or strongest earthquake to within a few-days interval and at a handful of locations globally*.

| ALIGNMENTS | DATE | MOON | LOCATION | MAGNITUDE | DEPTH | LABEL |
|---|---|---|---|---|---|---|
| Saturn-Earth-Sun | Nov 16 2000 | | Papua NG | Mw=8.0 | d=33.0 km | PS |
| Mars-Earth-Mercury; Earth-**Moon**-**Sun** | Jun 23 2001 | N+1 | Peru | Mw=8.4 | d=33.0 km | MPS |
| Earth-Venus-Mercury; Earth-**Moon**-**Sun** | Nov 03 2002 | N | Alaska US | Mw=7.9 | d=5.0 km | MPS |
| Earth-Moon-Sun | Sep 25 2003 | N | Japan | Mw=8.3 | d=27.0 km | MS |
| Earth-Mercury-Venus | Dec 23 2004 | | Macquarie NZ | Mw=8.1 | d=10.0 km | P |
| Earth-Mercury-Venus; Moon-Earth-Sun | Dec 26 2004 | F | Indonesia | Mw=9.1 | d=30.0 km | MPS |
| Earth-Mercury-Sun; Earth-Sun-Venus | Mar 28 2005 | | Indonesia | Mw=8.6 | d=30.0 km | PS |
| Jupiter-Earth-Sun | May 03 2006 | | Tonga | Mw=8.0 | d=55.0 km | PS |
| Earth-Sun-Venus-Jupiter | Nov 15 2006 | | Kuril Isles RU | Mw=8.3 | d=30.3 km | PS |
| Earth-Sun-Mercury | Jan 13 2007 | | Kuril Isles RU | Mw=8.1 | d=10.0 km | PS |
| Mars-Earth-Elenin; **Moon**-Earth-**Sun** | Apr 01 2007 | F | Solomon Isles | Mw=8.1 | d=10.0 km | CMPS |
| Earth-Venus-Elenin | Aug 15 2007 | | Peru | Mw=8.0 | d=39.0 km | CP |
| Earth-**Moon**-**Sun** | Sep 12 2007 | N | Indonesia | Mw=8.5 | d=30.0 km | MS |
| Earth-**Moon**-**Sun** | Sep 12 2007 | N | Indonesia | Mw=7.9 | d=30.0 km | MS |
| Earth-Moon-**Sun** | Dec 09 2007 | N | Fiji | Mw=7.8 | d=149.0 km | MS |
| Elenin-Earth-Neptune | May 12 2008 | | China | Mw=7.9 | d=19.0 km | CP |
| Earth-Mercury-Jupiter | Jan 03 2009 | | Indonesia | Mw=7.7 | d=17.0 km | P |
| Earth-Sun-Mercury | Jul 15 2009 | | New Zealand | Mw=7.8 | d=12.0 km | PS |
| Earth-Venus-Elenin | Sep 29 2009 | | Samoa | Mw=8.1 | d=18.0 km | CP |
| Venus-Earth-Uranus | Oct 07 2009 | | Vanuatu | Mw=7.8 | d=35.0 km | P |
| Elenin-Earth-Sun; Earth-Sun-Jupiter;MS | Feb 27 2010 | F | Chile | Mw=8.8 | d=23.0 km | CMPS |
| Elenin-Earth-Sun; Earth-Mercury-Uranus | Mar 11 2011 | | Japan | Mw=9.0 | d=32.0 km | CPS |

Table 2. Comparison of Earth's all long alignments at peak-times, v. very strong (~M8+) earthquakes of the decade of 2000's (expanded by earthquakes from year 2000 and the available (January-March) part of 2011). Labels as before.

Proximity of the Earth to alignments of other celestial bodies, particularly those involving the Sun, can result in intense seismic activity such as at the Yellowstone, US, coinciding with a swarm of thousands of quakes during the Earth's proximal pass by the Mars-Sun-Venus alignment of 17-25 Jan. (Similarly, proximity of the Sun to the many-planets alignment of 1 August 2010 coincided with the most massive solar-eruptions swarm enveloping half the Sun.) Even when coming in triplets such as of 21 Oct, short alignments seem unable to cause strong earthquakes. Note that the alignments are relatively rare events that occur on average less than once a week, while the multiple ones occur on average less than once a month.

Table 2 shows that the epochs of the very strong earthquakes from the decade of 2000's have coincided with alignments at around the peak-times. Note that, in terms of very strong seismicity, the Elenin started playing a role in 2007, and continued doing so, contributing in 6 out of the 30 alignment-relating seisms. Similarly, the Sun has participated in 19 such alignments, the Mercury in 9, the Venus in 8, the Moon in 9, the Mars in 2, the Jupiter in 4, the Saturn in 1, the Uranus in 2, and the Neptune in 1. The planetoid Pluto played no role.

Thus the planets generally play lesser role in seismicity-relating alignments the farther they are from the Sun, not from the Earth. Besides the fact that this can be expected (since orbital periods generally become longer with increase in orbital radius), this is also in agreement with the hyperresonator extension of the georesonator concept: here empirically demonstrated mechanism for the generation of seismicity lies outside the Earth. Note that alignments with the Sun can last up to three of more days due to the Sun's size and, consequently, a greater gravitational shadow.

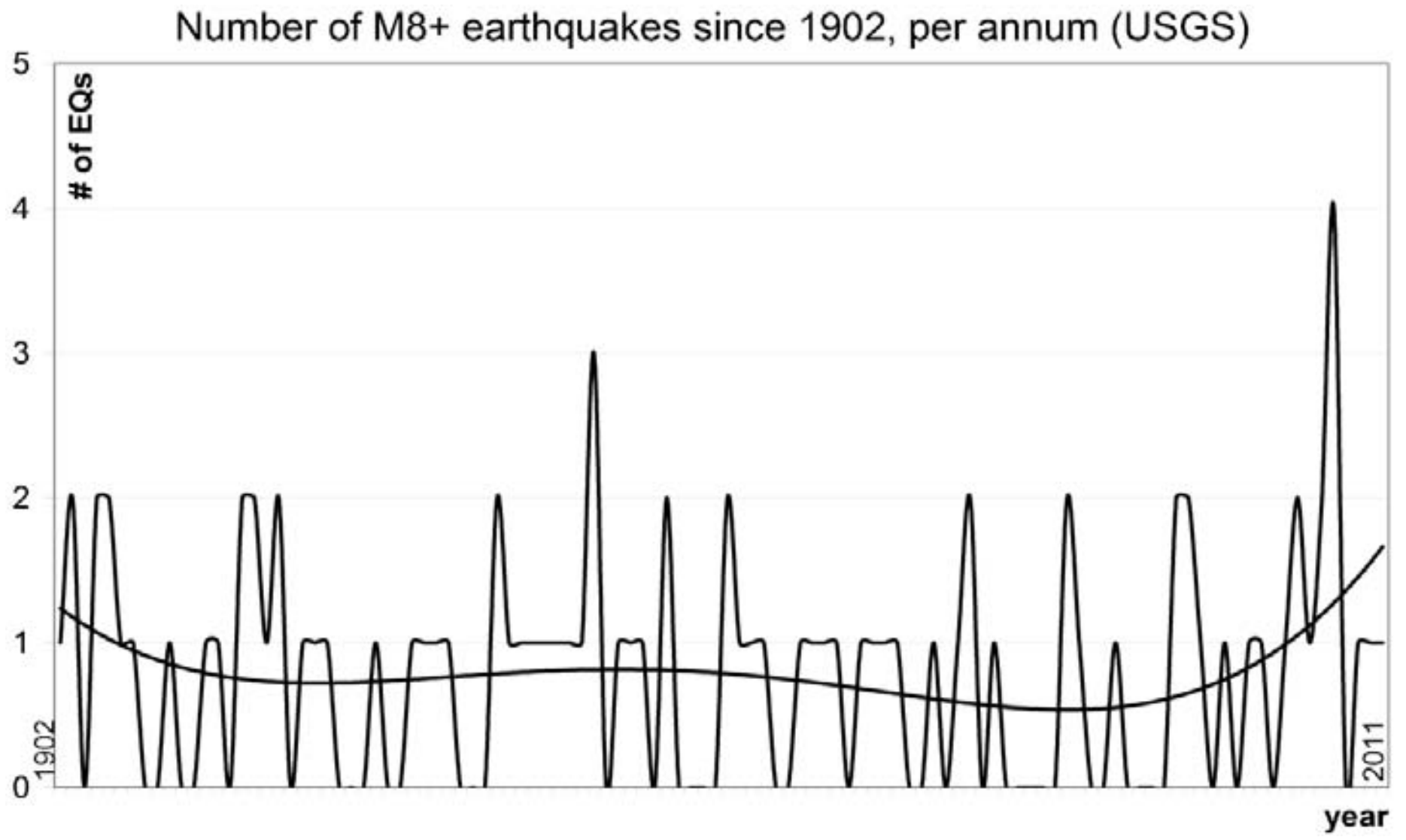

Figure 2. Occurrence of M8+ earthquakes 1902-2011 (so far), per annum. Note the largest increase in 2007, coinciding with the Elenin starting to play a role in M8+ seismicity-related alignments, Table 2. Polynomial trend of 4th order.

| ALIGNMENTS | DATE | Δ*t* | MOON | LOCATION | MAG. | DEPTH | LABEL |
|---|---|---|---|---|---|---|---|
| Earth-Sun-Mercury; Earth-Sun-Venus | May 22 1960 | 8,45 | | Chile | Mw=9.5 | d=33.0 km | PS |
| Earth-Mercury-Jupiter; Earth-Sun-Mercury | Mar 28 1964 | 12,24 | F | Alaska US | Mw=9.2 | d=23.0 km | MP |
| Earth-Mercury-Venus; Saturn-Earth-Sun; MS | Dec 26 2004 | 28,40 | F | Indonesia | Mw=9.1 | d=30.0 km | MPS |
| Elenin-Earth-Sun; Earth-Mercury-Uranus | Mar 11 2011 | 6, 8 | | Japan | Mw=9.0 | d=32.0 km | CPS |
| Jupiter-Earth-Sun; Jupiter-Earth-Mercury | Nov 04 1952 | 12,15 | | Kamchatka RU | Mw=9.0 | d=30.0 km | MPS |
| Elenin-Earth-Sun; Earth-Sun-Jupiter; MS | Feb 27 2010 | 7, 4 | F | Chile | Mw=8.8 | d=23.0 km | CMPS |
| Earth-Sun-Venus; Neptune-Earth-Uranus | Jan 31 1906 | 30,20 | | Ecuador | Mw=8.8 | d= unknown | PS |
| Elenin-Earth-Sun; Jupiter-Earth-Saturn; MS | Feb 04 1965 | 12,80 | N+3 | Alaska US | Mw=8.7 | d=Buldir block | CMPS |
| Earth-Mercury-Sun; Earth-Sun-Venus | Mar 28 2005 | 4, 7 | | Indonesia | Mw=8.6 | d=30.0 km | PS |
| Earth-Mercury-Saturn; Jupit-Earth-Sun; MS | Aug 15 1950 | 8,25 | N+2 | Assam, Tibet | Mw=8.6 | d=unknown | MPS |
| Earth-Mercury-Venus; Jupiter-Earth-Sun | Mar 09 1957 | 4,12 | | Alaska US | Mw=8.6 | d=33.0 km | P(S) |

Table 3. Comparison of the top 11 greatest (i.e., all strongest, M8.6+) earthquakes recorded instrumentally (since 1902), and the matching long alignments. The dashed line separates the high-strongest (M9+) earthquakes, and Δ*t* denotes estimated duration (in days) of the alignments. Labeling and luni-solar phases/eclipses are as in Table 1.

Table 3 shows eleven strongest (M8.6+) quakes recorded instrumentally (since 1902) v. the matching alignments. Multiple alignments can be noted on the day of the earthquake. Remarkably, three out of six (five, expanded arbitrarily by an M8.8 event) high-strongest (M9+) earthquakes recorded to date occurred during Full Moon. Four of the strongest recorded earthquakes have occurred in the last decade, and two in the last year due to the approaching Elenin.

The Sun participated in 18 of 26 long alignments, the Mercury in 9, the Venus in 5, the Moon in 4, the Jupiter in 7, the Elenin in 3, the Saturn in 3, the Uranus in 2, and the Neptune in 1. The Mars and the Pluto played no role. Thus a general trend as in Table 1 is largely preserved here too, even more so for strongest earthquakes, thus corroborating the georesonator concept in principle as well as empirically, Table 4 and Figure 4. These changes, which exhibit regularity, represent a most convincing indicator of celestial origin of strong seismicity on Earth.

Interestingly enough, the Mars has played no role in the occurrence of strongest earthquake examined, so it is likely that Mars contributes mostly to increase in ~M8.6– seismicity only. The Pluto has played no role in any of the earthquakes examined, so its recent demotion seems justified from the georesonator point of view as well: its orbital inclination of nearly 20° forbids the Pluto to partake in alignments, thereby corroborating the herein established alignments-seismicity causality. Importantly, all other planetoids from the Kuiper belt also have forbidding inclinations.

If luni-solar eclipses play a role, then it is only general (Table 1) albeit not particular, e.g., they seem to play no role at all in the high-strongest (M9+) seismicity. This in turn would mean that their general role is not real either, and that it appears as a role only because of a denser sample when more earthquakes are considered; given that there is 4-6 eclipses of any kind per year.

A phase of up to 3 days can occur in lunar forcing which causes deep earthquakes [2]. This is evident here for the Alaska, 1965 earthquake and, possibly, for the Tibet, 1950 earthquake too.

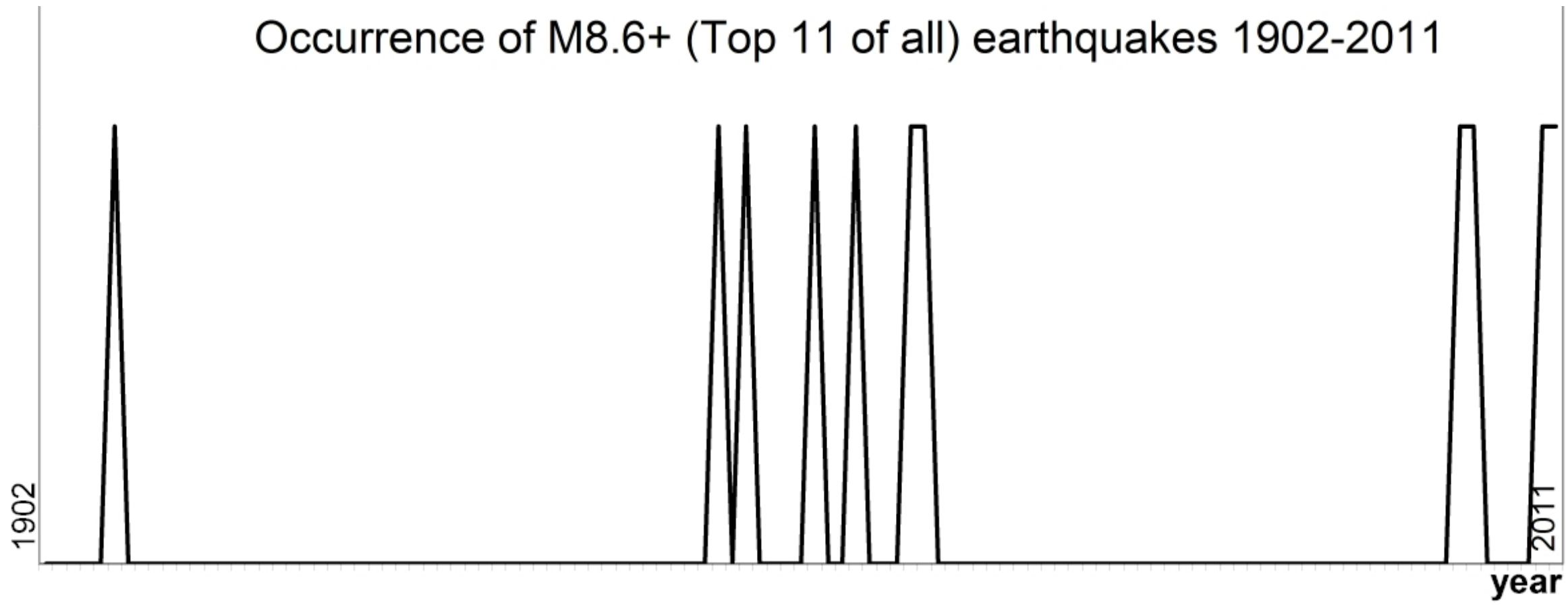

Figure 3. Occurrence of M8.6+ (the top 11) earthquakes between 1902 and March 2011, per annum. Note that these earthquakes came largely in subsequent pairs during the last century. This indicates intensity increase in georesonator's external forcing, coinciding with the start of the Elenin's role (in 1965) in seismicity-relating alignments, Table 3, or it suggests another solar system body from around that time that possibly played a role similar to that of the Elenin's.

| Of 30 long alignments with M8+ quakes of 2000's, Table 2, our solar system's bodies participated in: | | | | | | | | | | |
|---|---|---|---|---|---|---|---|---|---|---|
| Sun | Mercury | Venus | Moon | Mars | Elenin | Jupiter | Saturn | Uranus | Neptune | Pluto |
| 19 | 9 | 8 | 9 | 2 | 6 | 4 | 1 | 2 | 1 | 0 |
| Of 26 long alignments with M8.6+ quakes since 1902, Table 3, our solar system's bodies participated in: | | | | | | | | | | |
| Sun | Mercury | Venus | Moon | Mars | Elenin | Jupiter | Saturn | Uranus | Neptune | Pluto |
| 18 | 9 | 5 | 4 | 0 | 3 | 7 | 3 | 2 | 1 | 0 |

Table 4. Number of times a solar system body participated in Earth's long alignments under which M8+ of 2000's (top), and M8.6+ (bottom panel) earthquakes of the past 108 years occurred. The Elenin placed in the middle in both series.

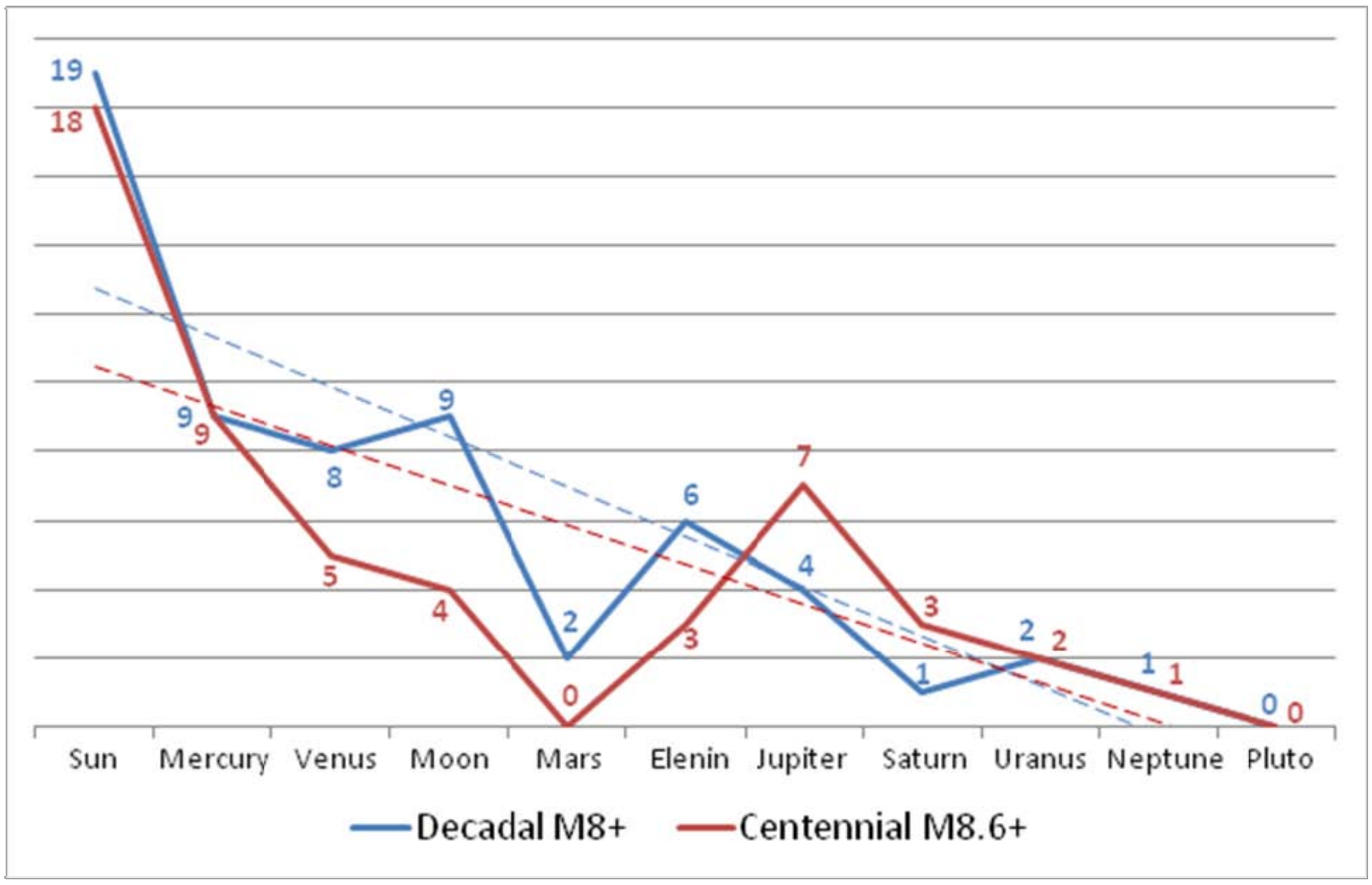

Figure 4. Regularity of change in the number of times a solar-system body participated in Earth's long alignments during which M8+ earthquakes of the decade of 2000's (blue), and M8.6+ (red) earthquakes of the past 108 years, have occurred, Table 4. The Elenin placed normally in the middle in both series. Remarkably, the Mars seems to be an "exception that confirms the rule", as the only planet in our solar system that plays virtually no role in Earth's strongest seismicity, in addition to the planetoid Pluto (but due to Pluto's large orbital inclination). Respective linear trends are dashed. Our Moon occasionally impacts very strong seismicity, same as other heavenly bodies in our solar system do. Note that, if the hyperresonator were not real, the values on this figure would be random and trends horizontal.

Note that oscillations add up, so the Earth's order within long alignment is of no significance for seismicity; meaning, whatever the Earth's position (conjunctional; oppositional) within the alignment, peculiarities considering the Earth seismicity inevitably arise in one form or another. This can be said even more of the Sun's position in its alignments involving the Earth, as the Sun is the central body to all possible alignments within its system so that the Sun partakes in most of Earth seismicity-relating alignments as well, Figure 4. Given such a crucial role that the Sun plays in Earth seismicity-related alignments, its effects should also be alternating in a more drastic fashion when it is in opposition versus conjunction. Indeed, when the Sun is in opposition and combined with Full Moon (only highlighting the hyperresonance effect), extraordinary swarms like that between 27.02-01.03.2010 occur with some 300 M5+ ($M_w$ and $M_b$) earthquakes under the Elenin-Earth-Sun alignment; similarly between 21-22.12.2010 with some 100 M5+ ($M_w$ and $M_b$) earthquakes under the Earth-Mercury-Sun alignment. On the other hand, when the Sun is in conjunction and combined with Full Moon, complete shutdown of M5+ seismicity occurs, Figure 14. Similarly, same effect of seismicity shutdown is seen for the Sun in opposition but combined with New Moon, Figure 15.

Obviously, the largest absence of land mass on the georesonator will necessarily act as a natural attenuator. This in turn explains relative seismic tranquility of inland v. oceanic regions in terms of very strong earthquakes. It is thus evident why the Pacific Ocean, as the largest landmass discontinuity, makes for the most suitable region for hosting the Ring of Fire. The Earth simply absorbs most of its magnified vibration by locking up the wave energy that naturally moves from all directions towards the Pacific Rim as the largest landless area of all. The largest area naturally picks up the most of the vibration, and even more so when a vibration magnification is concerned. By acting as the attenuator, the circum-Pacific makes other similar regions relatively calm, such as the circum-Atlantic or the circum-Mediterranean (which together make the second largest body of mass, and thereby overall the second-most shaking region on Earth, too). As a long alignment nears, mechanical foreshocks occur normally in a matter of days to weeks prior to the main shock, followed by mechanical aftershocks as the alignment fades out. Besides, the only all-encompassing explanation for continuous periodicity in the occurrence of strong earthquakes, as well as for their occurrence within tectonically inactive regions, lies in long astronomical alignments as the only truly periodic events on the centennial scale. What some have claimed to be years-delayed aftershocks are in fact series of strong earthquakes arising seemingly separately but still due to same or similar astronomical conditions. Then what we usually refer to as tectonic/seismic faults, subduction zones, trenches, etc. from the theory of plate tectonics, are merely material-aging fractures on the georesonator, resembling cracks on a bridge as soldiers step march across it and giving rise to the bridge's resonance magnification.

Accurately determining the location of strongest earthquakes is a matter of applying the existing structural engineering knowledge to future data gathered via globally gridded networks of sensors on global scales. Given that the demonstration of georesonator also proves the hyperresonator, it can be said that, generally, a larger gravitational shadow will result in a more energetic response by the georesonator, albeit not proportionately to either the body's mass or distance from the Earth (but from the Sun instead, Figure 4). I showed earlier that the dependence is on the body's eigenfrequencies i.e. apparent size [3]. Obviously, the georesonator is the exact solution to a specific case of the previously unsolved three-body problem of the Newtonian mechanics.

The increase in seismic activity due to combined effects of the Elenin and other celestial objects is manifested in form of phases longer than 3 days, and displacements of more than 10 m. As predicted in versions 1-4 of the arxiv.org preprint of this paper, the increase has indeed continued until the comet crossed into the Earth's orbit around 1 August 2011, when the seismicity due to Elenin subsided as this comet sped up significantly. After that, the Elenin did not have enough time to participate in many long alignments in order to pose imminent danger to the Earth as before. The Elenin regained its role (see Appendices) of a major factor in Earth's seismicity past 20 October 2011, i.e. upon the comet's crossing of the Earth's orbit on the way out, which will continue until subsiding around 2016. Since it survived the September, 2011, rendezvous with the Sun, the Elenin's role will finally fade out in terms of very strong seismicity within a few decades.

The georesonator concept explains why atmospheric data behave as seismic precursors: the atmosphere is just a part of the masses that get thrown into oscillation magnification together with the mantle, rocks and oceans. Therefore the atmosphere reacts along the vector of the alignment, in a precursory fashion. But such a behavior is just a byproduct of astronomical forcing of the georesonator, and as such it arises from the same (external) sources as seismicity does. Atmospheric and other precursors are not causally related to earthquake processes inside the Earth, but belong simultaneously to a same process of outer forcing of all the Earth masses. Thus atmospheric precursors do not consist only from gas emission, which some allege extends upwardly from the Earths' depths to the atmosphere. For example, it has been established that GPS soundings of the ionosphere show distinct changes days ahead of some very strong earthquakes. Therefore, it is the mechanical shaking via magnification of mass resonance, and not the seismic stress release via rock compression, which is the prime cause for the release of gases days ahead of activation of the seismic fault itself. (Simply, stress release cannot take place over such relatively large areas as it was shown by others.) Here 'prime' is by extension: since mass-resonance magnification is the prime energy supplier and thus the cause of strong earthquakes too, it is also the prime cause of their precursors. Similarly, the ionosphere disturbances are just a part of the same mechanism of resonance magnification.

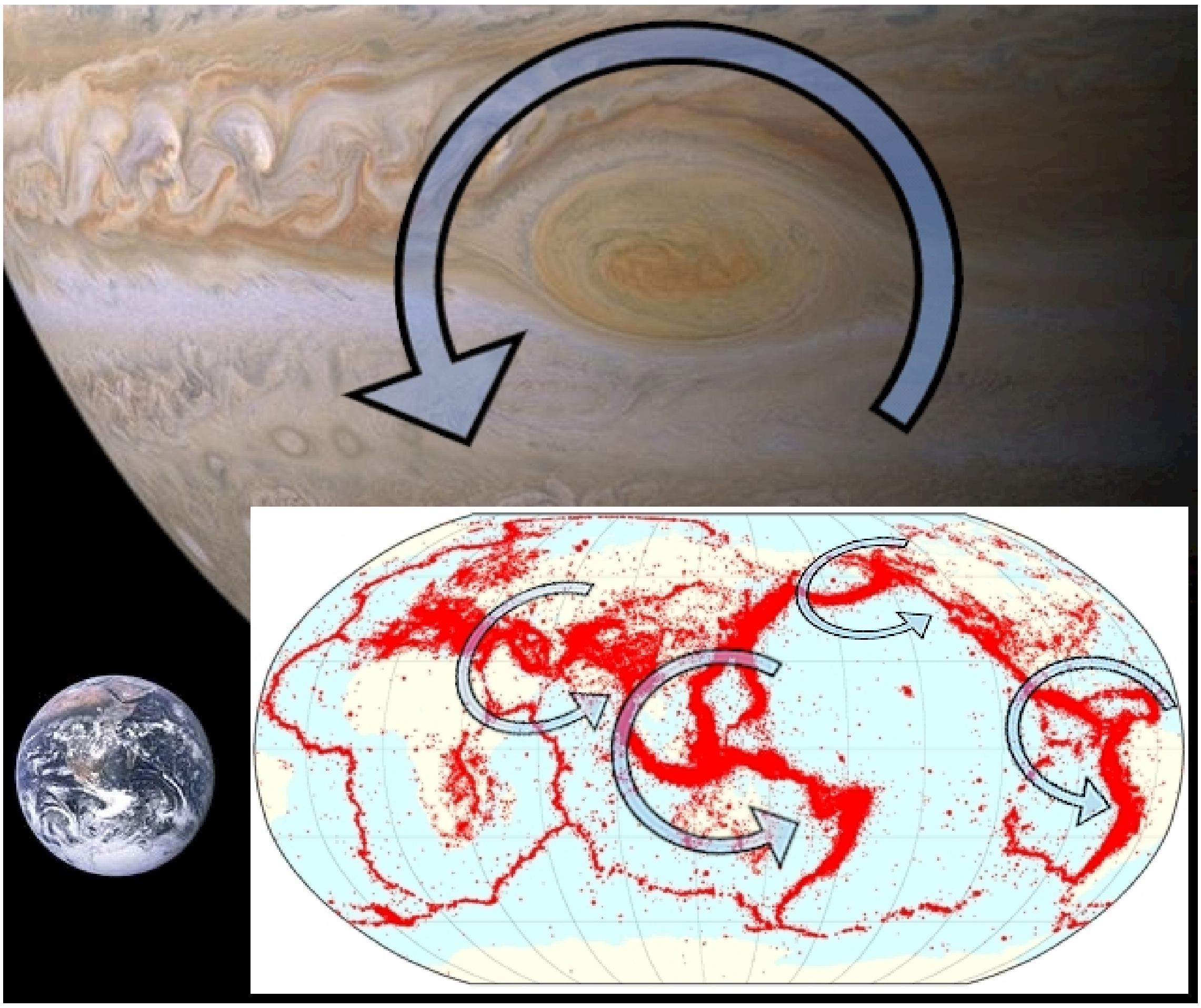

Figure 5. Comparison of the Jovian Great Red Spot and the Earth's most seismic zones. Both phenomena maintain same (counter-clockwise) rotational preference over long geological times, regardless of what hemisphere they are in, unlike with domestically caused mass-vortex phenomena such as cyclones and anti-cyclones. Since it is not necessarily of domestic origin, and is seen on both planets (i.e., regardless of mass density), the phenomenon is caused externally – by magnified mass resonance due to our Solar system as the only candidate. Images of the Jupiter v. Earth to scale. Map modified from: World Seismicity Map 1964-2008, International Seismological Centre. Robinson compromising map projection, arguably the most faithful representation of the Earth surface everywhere except in frigid zones, used here to demonstrate the hyperresonance based on the facts that: (i) all the largest strain-vortices on Earth visibly obey the georesonator principle, and (ii) rotational direction of strain-vortices is preferential i.e. regardless of the hemisphere.

Then, if the atmosphere (gas) were of significantly firmer viscosity, we would be experiencing strong atmospheric seismicity too [2]. Inversely then, if the Earth were gaseous instead of solid, its vortices of inner-structural failure as due to georesonance, such as in Oceania or South America or Euro-Asia or Alaska, would still be continuously maintained externally by the resonance magnification, except the vortices would then appear in the same manner and with always the same rotational direction (regardless of the hemisphere) as the Jovian Great Red Spot does due to magnified mass-resonance, or *jovioresonance*; Figure 5. The rotational preference as seen on Figure 5 is due to the direction of planetary revolution about the Sun in our solar system, where planets have tendency to behave rather like pinwheel toys in the wind.

The herein given empirical proofs, together with previous observational and theoretical proofs, demonstrate beyond doubt that the alignments do cause (via hyperresonance mechanism) strongest, very strong, and strong earthquakes. Examination into a relationship between alignments and strong seismicity served the purpose of demonstrating the georesonator's mechanical principle via finding the expected path (increasing-peaking-decreasing) of magnified oscillation during the alignment. At the same time, examination into a relationship between alignments and very strong seismicity makes the core of this paper. Note that here established causality is by exclusion: it stems from the fact that there are no alternative intermediaries in Space (relating earthquakes and alignments indirectly), other than the hyperresonator [3]. That is, there exists no physical entity which could act as an intermediary relating the whole Earth's energy emissions to an (any!) astronomical phenomena whatsoever. So the correlation is real.

Very long alignments (on the order of weeks-to-months) also exhibit a regular magnitude trending: as one resonance magnification dies out under a very long alignment, another one begins, with or without a delay. This is fortunate because if mass resonance were not dying out naturally, the Earth would have been destroyed by resonance magnification a long time ago. Very long alignments show nicely that mass-resonance magnification outputs enough energy to stir things up, but insufficient to destroy an object, except in special cases like fragile construction – the 1940 example of the collapse of Tacoma bridge in the US. Also, peak-magnitudes of each magnification episode in a very long alignment themselves obey in a fractals' fashion the same pattern as the strong earthquakes' magnitudes within a long alignment do (see Appendices).

Note finally that strong seismicity during long alignments has mostly to do with the object's orbital inclination (which must be virtually coplanar to our solar system for any alignment to occur), relatively low speed, and its en-masse relatively large size. Such sizeable, non-periodic comets, and which have a hyperbolic trajectory that is virtually coplanar to our solar system, as is the case with the Elenin, are very rare and come at a rate of about a few per century. That is what had made the Elenin so useful for this study. Then just as with the planets, the comet's inclination is also the primary factor which controls the comet's role in the seismicity of aligned bodies. Since the georesonator concept is universal, it should apply to other objects as well; for example, the Sun's largest coronal mass ejection ever recorded – that of 7 June 2011 – had occurred during multiple alignments Sun-Venus-Mars and Elenin-Sun-Uranus. This was a Sun's equivalent of a M9+ earthquake due to the Elenin.

## Conclusions

The Earth's strong seismicity is unrelated to tectonics, and they both arise due to the same external (astronomical) causes, namely the Earth's position with respect to the solar system's constellations. So the long alleged mutual causality of strong earthquakes and tectonics is not real either. ~M6- seismicity is primarily due to tectonics and, secondarily, to the mantle's resonance magnification, since the mantle's permanent dynamics is at play, caused all the time by the resonance magnification from some alignments.

This discovery has significant effects in preserving human life and habitat, because days without strong earthquakes can now be known exactly. They amount to months per year of absolutely known seismic forecast which tells that there will be no catastrophic earthquake on that-and-that date. Besides, given the fact that the mechanism behind strong seismicity is now known, approximate forecast of strength, time and location of strong earthquakes for the rest of the year is readily available too. Combined with careful studies of the structural characteristics of the lithosphere, knowledge unveiled via the georesonator concept offers for the first time a real hope for accurate scientific earthquake forecast too. So an approximate forecast is already possible in magnitude, time and location. Magnitude-wise, this can be done within a few degrees of magnitude, as earthquakes shown here to be caused by alignments are of the ~M6+ strength, while multiple alignments cause ~M7+ quakes. Time-wise, this can be narrowed down to within a few days by monitoring the Earth's response to an alignment as it occurs, once it occurs. The location too can be narrowed down once the alignment commences, as well as by establishing which exact alignments are responsible for causing very strong seismicity in those areas that are known to shake in cycles on century scales, such as the San Francisco Bay area. Thus the location can also be narrowed down to the most probable geographic frontier of the regions/plates whose grave oscillation mode is expected to meet the collapse state considerably sooner than any others', as the gravitational shadow traverses the Earth (the candidate-regions). That this is physically justified stems from the seemingly odd fact that strong seismicity on Earth is rarest in the frigid zones.

This empirical demonstration, here finalized by the conclusive empirical proof showing a 6-months-long, 100%-accurate prediction of resonance ~M6+ magnification (Figures 14, 15, 17, 18; Appendices D, E), closes the three-step proof of *georesonator* [2] and its generalization *hyperresonator* [3], which concepts are hereby deemed proven. For the purpose of completion, I here simply reproduce my hyperresonance equations: [3]

$$\left| G_{Newton} \right|_{NIST} = (6.6742 \pm 0{,}001) \cdot 10^{-11}$$

|2006

$$\left| G_{Planck} \right|_{NIST} = (6.7087 \pm 0{,}001) \cdot 10^{-39}$$

$w_{\text{o Earth}} \,/\, w_{\text{Earth-Moon}}$ = 14.7655 day / 3455 sec = $C$

$$\left| c^{-1} \right| \cdot C^{-1} \cdot e^2 = 6.6750 \cdot 10^{-11} = \left| G_{Newton} \right|^*$$

$$\left| c^{-4} \right| \cdot C^{-2} \cdot e^2 = 6.7093 \cdot 10^{-39} = \left| G_{Planck} \right|^*$$

Or, written differently:

$$\left| G_{Newton} \right|^* = \left( \frac{T_{Earth}}{T_{Earth-Moon}} \right)^1 \cdot \frac{e^2}{|c|} = 6.6750 \cdot 10^{-11} = \left| G_{Newton} \right|_{NIST}$$

$$\left| G_{Planck} \right|^* = \left( \frac{T_{Earth}}{T_{Earth-Moon}} \right)^2 \cdot \frac{e^2}{|c|^4} = 6.7093 \cdot 10^{-39} = \left| G_{Planck} \right|_{NIST}$$

Note that the above, 2006 estimates of NIST constants improve the estimate of $T_{\text{o Earth}}$, from a coarse 3455 s based on previous estimate, to 3454.8 ± 0.1 s. And while the 2006 update closes the relationship an order of magnitude better than previous values used to, note also that the 2010 update scores against the 2006 update less than half an order of magnitude better.

Or, written in a short form, for a scale, $s$:

$$G = s \cdot e^2 ,$$

which obviously means the 4th Kepler Law.

The hyperresonator principle, as the main discovery of this 3-step proof, is akin to tuning forks triggering each other at a distance. This conclusion stems from the most obvious physical meaning of the hyperresonator equations: although the equations were derived for the Earth-Moon system, it is obvious due to their accuracy alone that the (any) body's grave mode of oscillation is not a random number as previously believed. Note here that the hyperresonator equations are accurate to at least 39 decimal places.

My discovery demonstrates the existence of aether as deduced in [3] and required for propagation of any (and thereby also the georesonator's) mutually tuning (interfering) oscillations. Gravity can now be expressed in terms of speed of light, i.e., true gravity is not a classical force but a totality of all magnified oscillations of the aether (*stringdom*). Gravity has two main components: apparent and magnified, the latter referred to as "dark energy/matter". The fork-tuning principle is then universal. The uniformity (both direction-wise and size-wise) of stringdom's strain-vortices as observed for our Solar system proves the hyperresonance empirically. Then other phenomena, such as spiral galaxies, "black holes" and "gravitational lenses" are strain-vortices ("pinwheel toys") of stringdom too.

# APPENDIX A

## Examples of alignments related to strongest earthquakes of the decade of 2000's

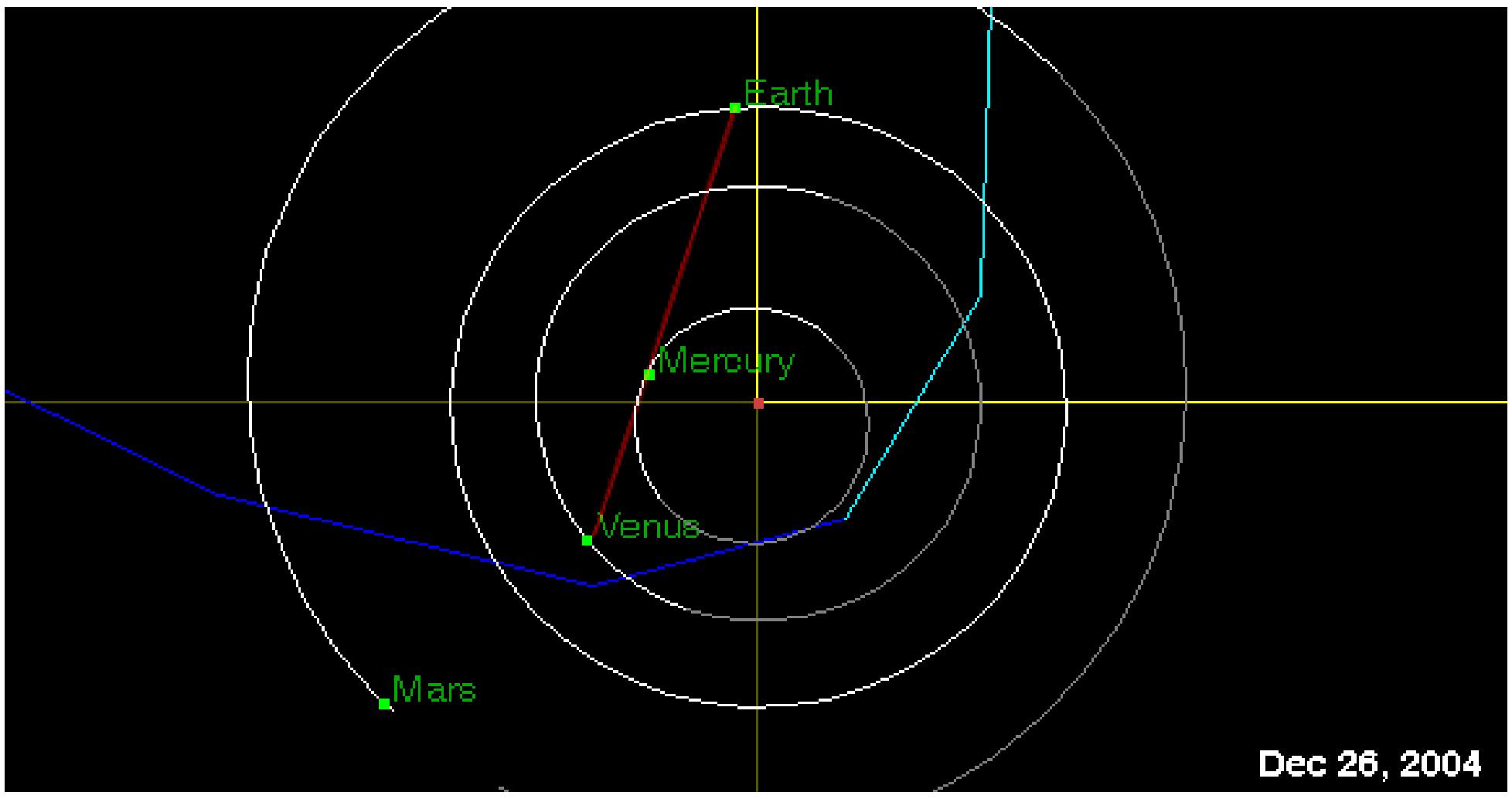

Figure 6a. Long alignment (red) Earth-Mercury-Venus on day of M9.1 Indonesia quake of 26-Dec-2004, Table 2.

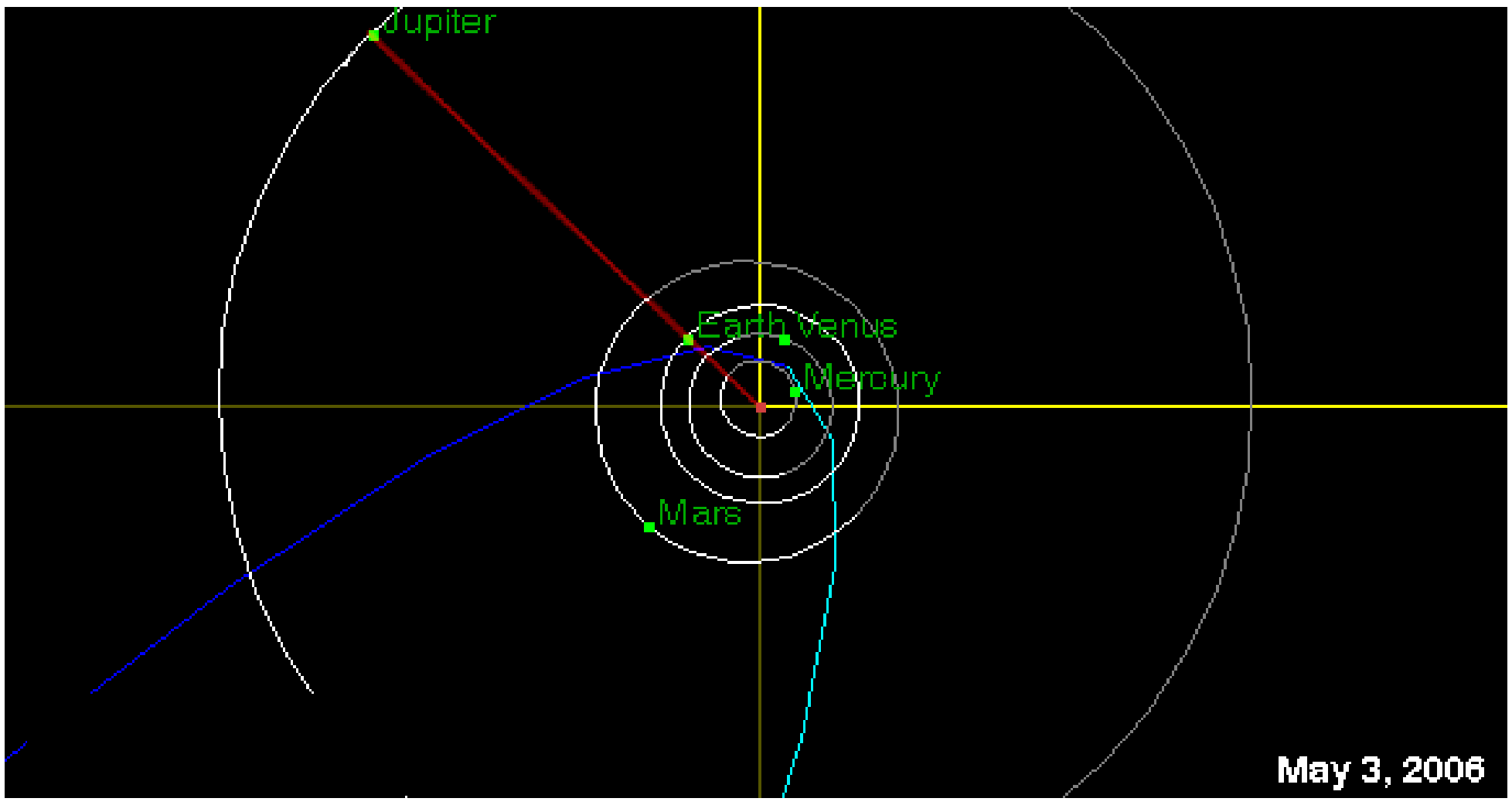

Figure 6b. Long alignment (red) Jupiter-Earth-Sun on the day of M8 Tonga earthquake of 03-May-2006, Table 2.

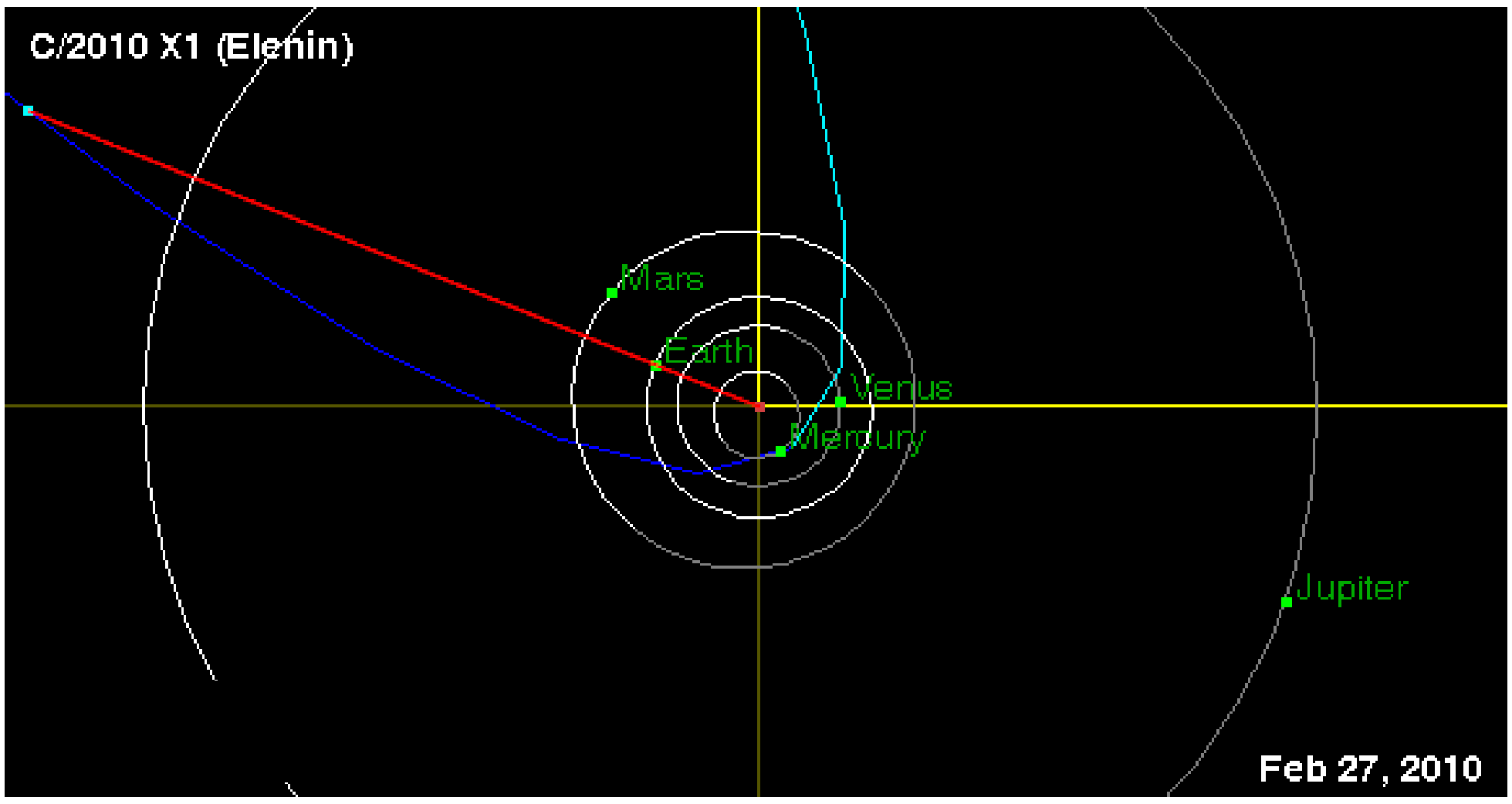

Figure 7. Long alignment (red) Elenin-Earth-Sun on the day of M8.8 Chile earthquake of 27-Feb-2010, Table 2.

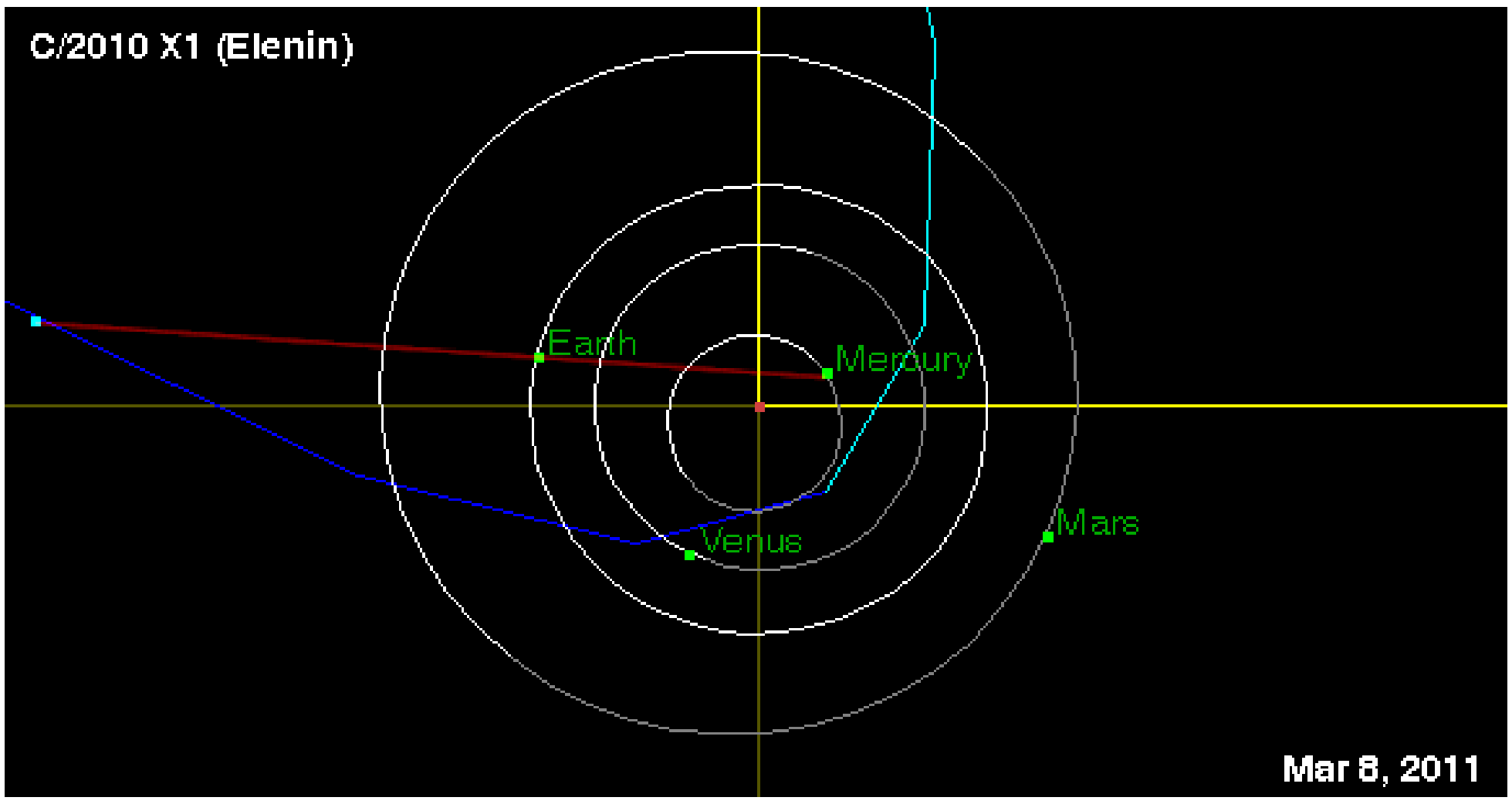

Figure 8. Long alignment (red) Elenin-Earth-Mercury, -3 day of M9 Japan earthquake of 11-Mar-2011, Table 2.

# APPENDIX B

**Example of continued mass resonance magnification (alignments-earthquakes relationship) tracked in near real time from May-July 2011**

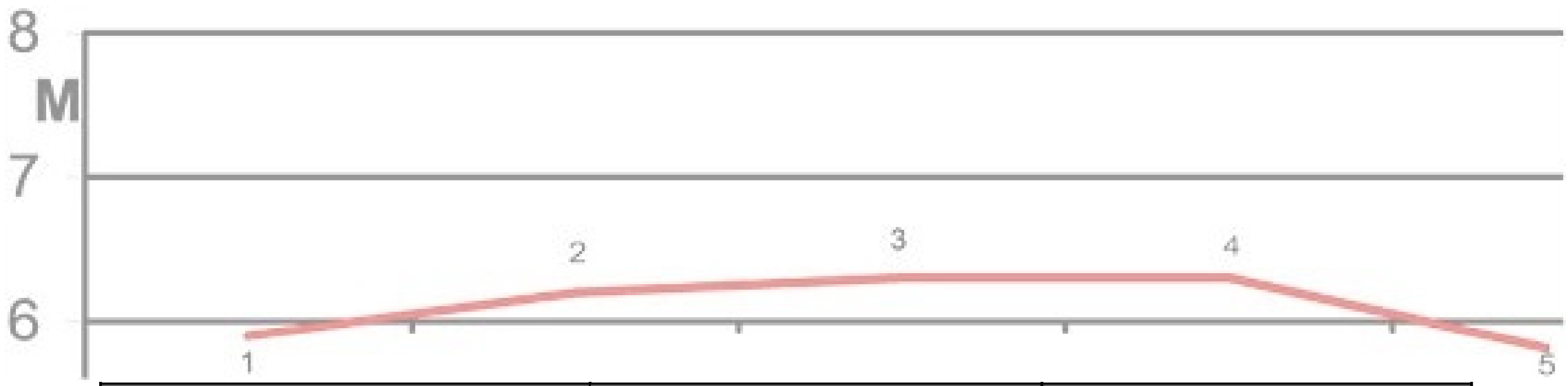

| | | |
|---|---|---|
| 1: 29.05 INDONESIA, M5.9 | 2: 01.06 BIO-BIO CHILE, M6.2 | 3: 03.06 HONSHU JAPAN, M6.3 |
| 4: 05.06 MACQUARIE ISLE, M6.3 | 5: 05.06 FIJI, M5.8 | |

Figure 9. Long alignment Elenin-Earth-Neptune (26 May-05 June).
Pattern of strong seismicity intensification: as expected.

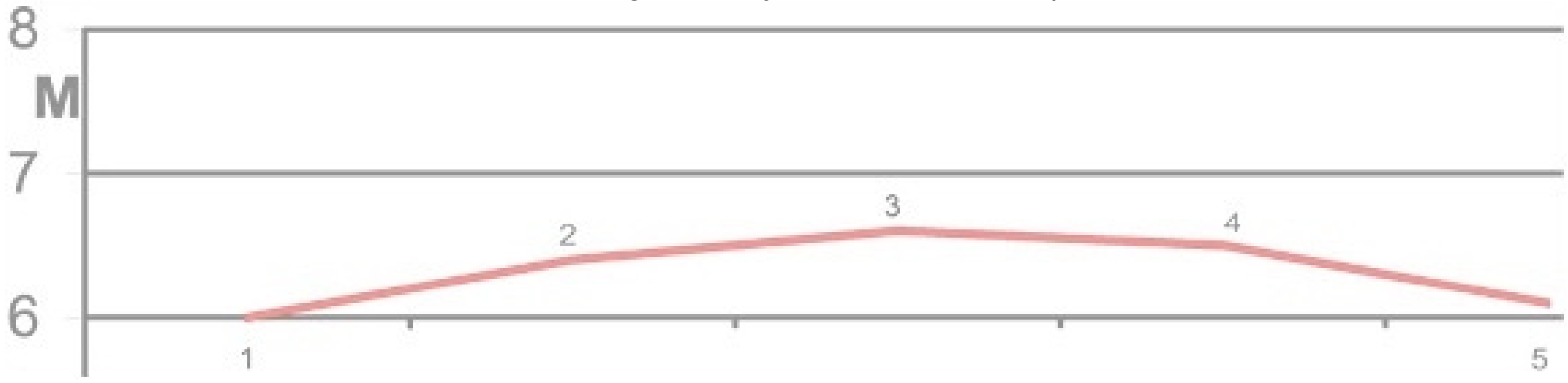

| | | |
|---|---|---|
| 1: 13.06 S.ISLAND N.ZEALAND, M6.0 | 2: 13.06 INDONESIA, M6.4 | 3: 16.06 PAPUA N.GUINEA, M6.6 |
| 4: 20.06 CHILE, M6.5 | 5: 21.06 STA CRUZ ISLES M6.1 | |

Figure 10. Long alignment Earth-Sun-Mercury (12-21 June).
Pattern of strong seismicity intensification: as expected.

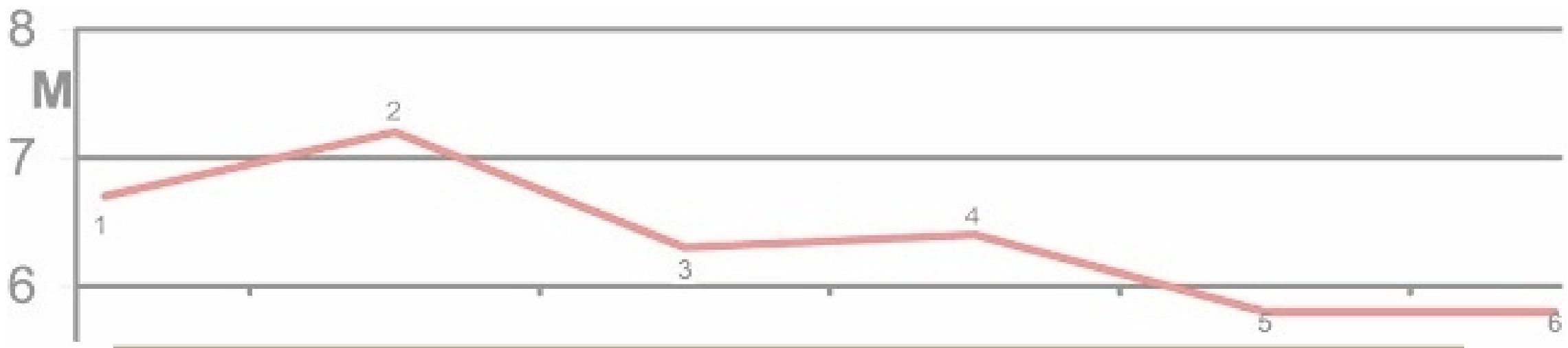

| | | |
|---|---|---|
| 1: 22.06 HONSHU JAPAN, M6.7 | 2: 24.06 ALASKA, M7.2 | 3: 24.06 STA CRUZ ISLES, M6.3 |
| 4: 26.06 INDONESIA, M6.4 | 5: 27.06 INDONESIA, M5.8 | 6: 28.06 ALASKA, M5.8 |

Figure 11. Long alignment Elenin-Earth-Neptune (22-30 June), as a continuation of the resonance magnification which was demonstrated for 26 May-05 June, so half of the ~M6+ earthquakes above were in the same regions as back then (top panel). Pattern of strong seismicity intensification: as expected.

Note that all of the intervals between the ones during which the above depicted earthquakes occurred, were quiescent in terms of strong seismicity: 20-25 May, 06-11 June, 01-04 July, as expected. Here quiescent means "within prescribed accuracy of up to one M6+ earthquake per interval of time spanning three or more days without Earth's long alignment".

The tracking was done at: www.sites.google.com/site/omerbashich/blog, as well as at: www.seismo.info. Note that status accuracy can improve only once the Earth goes into its long alignment. For the pattern enhancement purposes, the M(6-5%)+ (i.e. M5.7+) earthquakes were used, at the beginning and ending of a mass resonance magnification.

## APPENDIX C

**Conclusive proof of long alignments-seismicity relationship as arising from mass resonance magnification, tracked in near real time from May-July 2011**

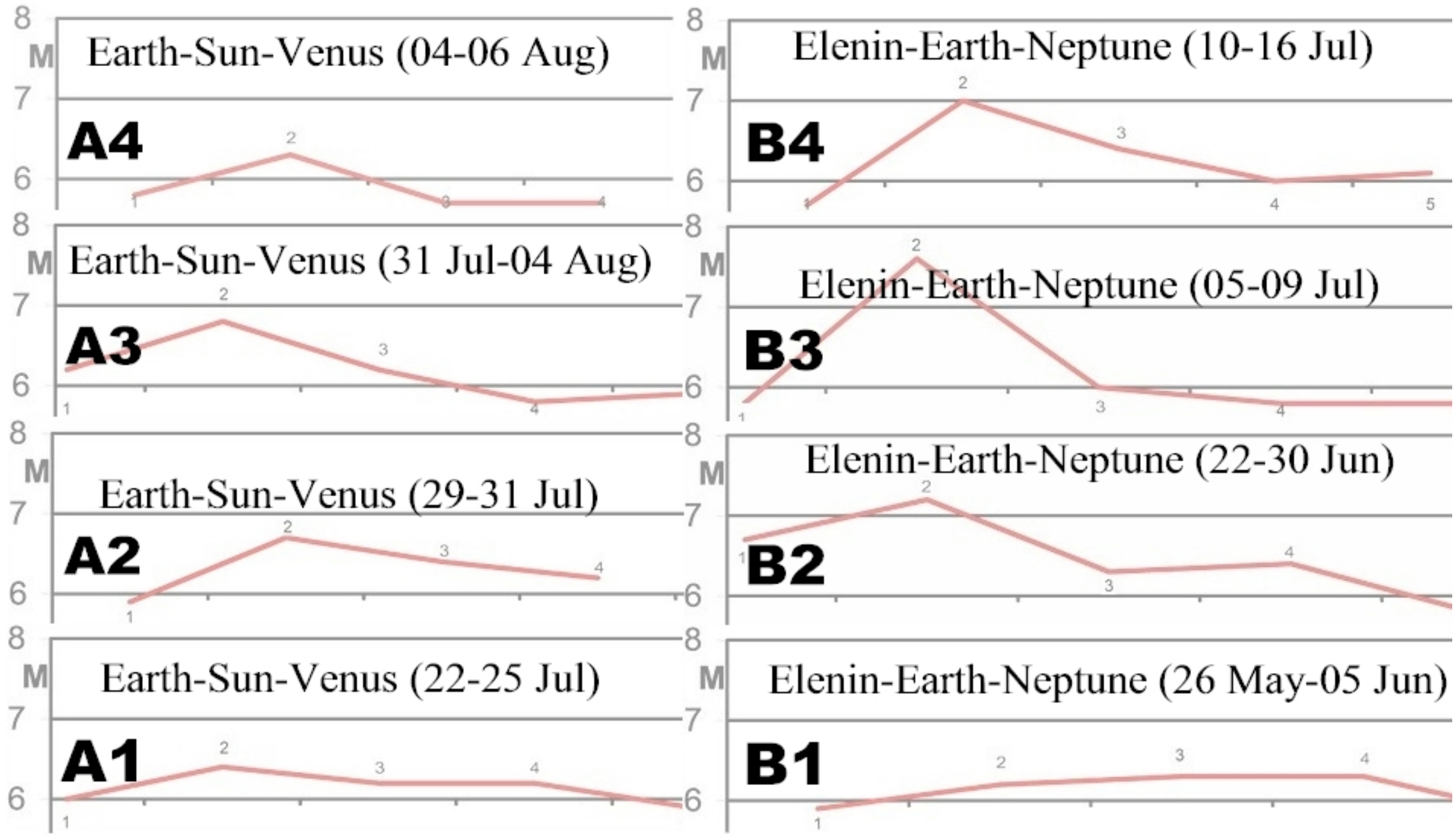

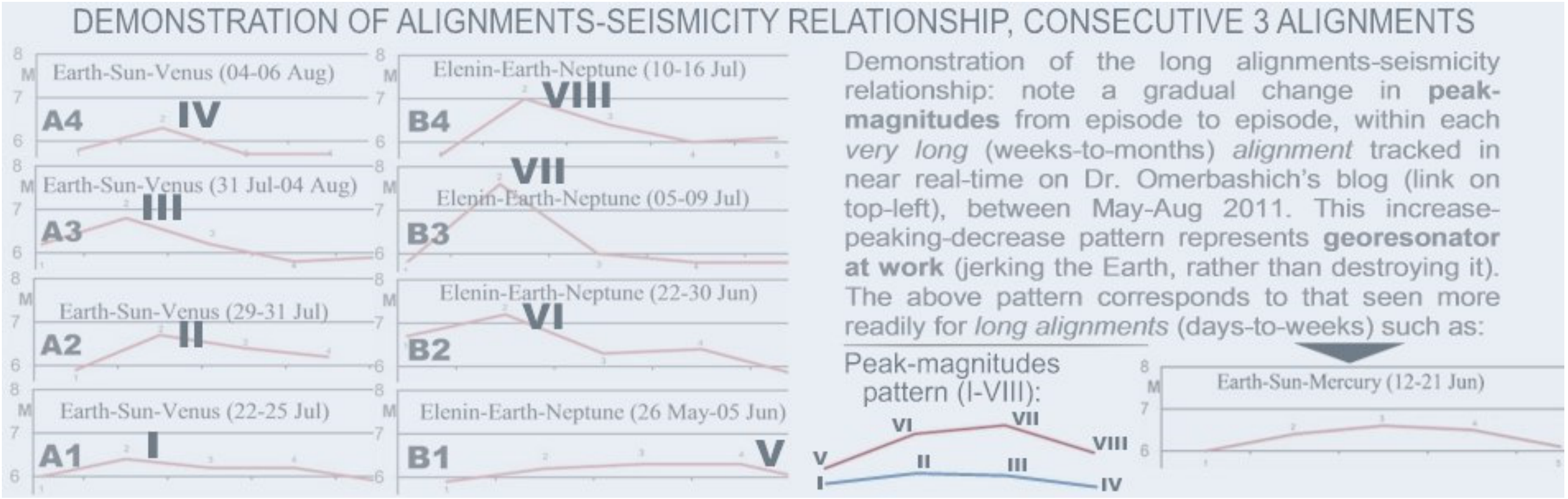

Figure 12: Demonstration of the long alignments-seismicity (Table 5) relationship: note a gradual change in peak-magnitudes, from episode to episode within each of the *very long alignments* ("on the order of weeks-to-months") tracked online in near real-time during May-August 2011. This increase-peaking-decrease pattern (red and blue, bottom panel) I-IV and V-VIII represents the **georesonator at work** (jerking the Earth, rather than destroying it). The above pattern corresponds to that seen more readily for *long alignments* ("on the order of days-to-weeks"), see plot in the bottom-right corner, bottom panel. In the above, M5.7+ earthquakes were used for enhancing the pattern at the edges only (beginning or/and end of mass-resonance magnification).

| | |
|---|---|
| A1.1: 22.07 FIJI REGION, M6.0<br>A1.2: 23.07 HONSHU JAPAN, M6.4<br>A1.3: 24.07 HONSHU JAPAN, M6.2<br>A1.4: 25.07 PAPUA N. GUINEA, M6.2<br>A1.5: 25.07 PHILIPPINES, M5.9<br><br>A2.1: 27.07 MID-ATLANTIC, M5.9<br>A2.2: 29.07 FIJI ISLES, M6.7<br>A2.3: 30.07 HONSHU JAPAN, M6.4<br>A2.4: 31.07 VANUATU ISLES, M6.2<br><br>A3.1: 31.07 VANUATU ISLES, M6.2<br>A3.2: 31.07 PAPUA N.GUINEA, M6.8<br>A3.3: 01.08 HONSHU JAPAN, M6.2<br>A3.4: 02.08 FIJI ISLES, M5.8<br><br>A4.1: 04.08 INDONESIA M5.8<br>A4.2: 04.08 KURIL ISL.RUSSIA M6.3<br>A4.3: 06.08 INDONESIA M5.7<br>A4.4: 10.08 PAKISTAN M5.7 | B1.1: 29.05 INDONESIA, M5.9<br>B1.2: 01.06 BIO-BIO CHILE, M6.2<br>B1.3: 03.06 HONSHU JAPAN, M6.3<br>B1.4: 05.06 MACQUARIE ISLE, M6.3<br>B1.5: 05.06 FIJI, M5.8<br><br>B2.1: 22.06 HONSHU JAPAN, M6.7<br>B2.2: 24.06 ALASKA, M7.2<br>B2.3: 24.06 STA CRUZ ISLS, M6.3<br>B2.4: 26.06 INDONESIA, M6.4<br>B2.5: 27.06 INDONESIA, M5.8<br><br>B3.1: 06.07 HONSHU JAPAN, M5.7<br>B3.2: 06.07 NEW ZEALAND, M7.6<br>B3.3: 07.07 NEW ZEALAND, M6.0<br>B3.4: 09.07 TONGA, M5.8<br><br>B4.1: 09.07 NEW ZEALAND, M5.7<br>B4.2: 10.07 HONSHU JAPAN M7.0<br>B4.3: 11.07 PHILIPPINES, M6.4<br>B4.4: 16.07 CHILE. M6.0<br>B4.5: 16.07 ALASKA, M6.1 |

Table 5. Data for Figure 12, USGS.

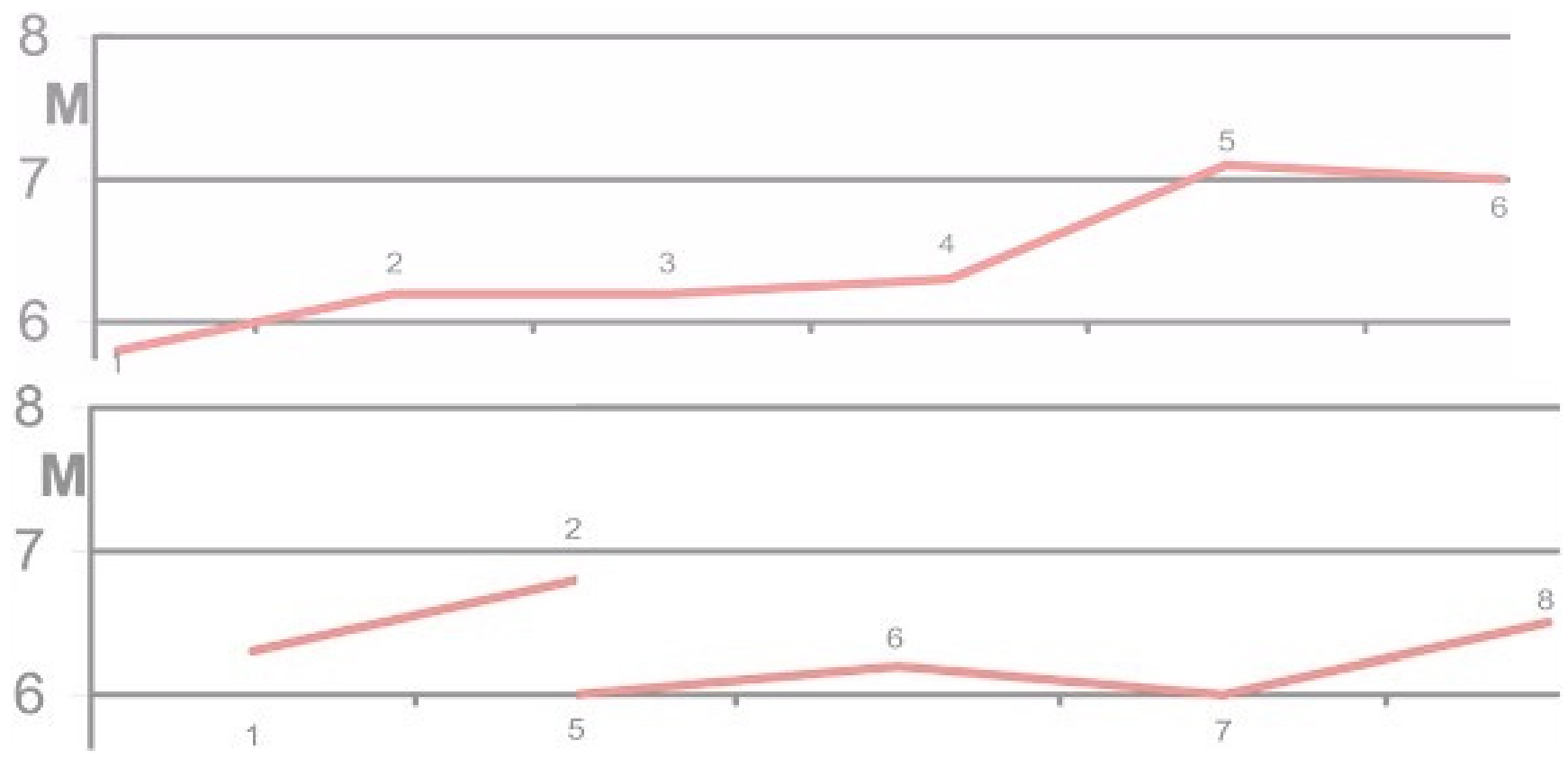

Figure 13: Fifth episode (top panel) of the very long alignment Earth-Sun-Venus (22 July and through mid-September), interposed by the Mercury during mid-two weeks of August. This interpose has introduced oscillatory-orbital interference which shut down the M6+ seismicity (Table 6) on the Earth during those two weeks. The mode of reactivation after the interposing (top) was then same as that after the previous interpose of the alignment Earth-Mercury-Venus of May 2011, by the Jupiter (bottom panel). Note that all time intervals in between the here demonstrated activity (in the sense of M6+ seismicity) were quiescent in the same sense, meaning no M6 earthquakes occurred in those intervals, e.g., the Earth was indeed quiescent in the same sense during the two-weeks interpose, only to see its mass-resonance magnification rebounding immediately after the interposing.

1: 16.08 INDONESIA M5.8
2: 17.08 HONSHU JAPAN, M6.2
3: 19.08 FIJI, M6.2
4: 19.08 HONSHU JAPAN, M6.3
5: 20.08 VANUATU, M7.1
6: 20.08 VANUATU, M7.0

1: 09.05 MACQUARIE ISLES, M6.3
2: 10.05 NEW CALEDONIA, M6.8
3: empty placeholder marking an interpose day 11.05
4: empty placeholder marking an interpose day 12.05
5: 13.05 COSTA RICA, M6.0
6: 13.05 HONSHU JAPAN, M6.2
7: 15.05 MID-ATLANTIC, M6.0
8: 15.05 PAPUA N GUINEA, M6.5

Table 6. Data for Figure 13 (top and bottom portions correspond to respective panels), USGS.

# APPENDIX D

**Demonstration of 100% accurate prediction of Earth's M6+ magnified oscillation due to long astronomical alignments, based on georesonator concept from 60 days of successive real-time tracking of the prediction in Oct-Nov 2011** (www.seismo.info)

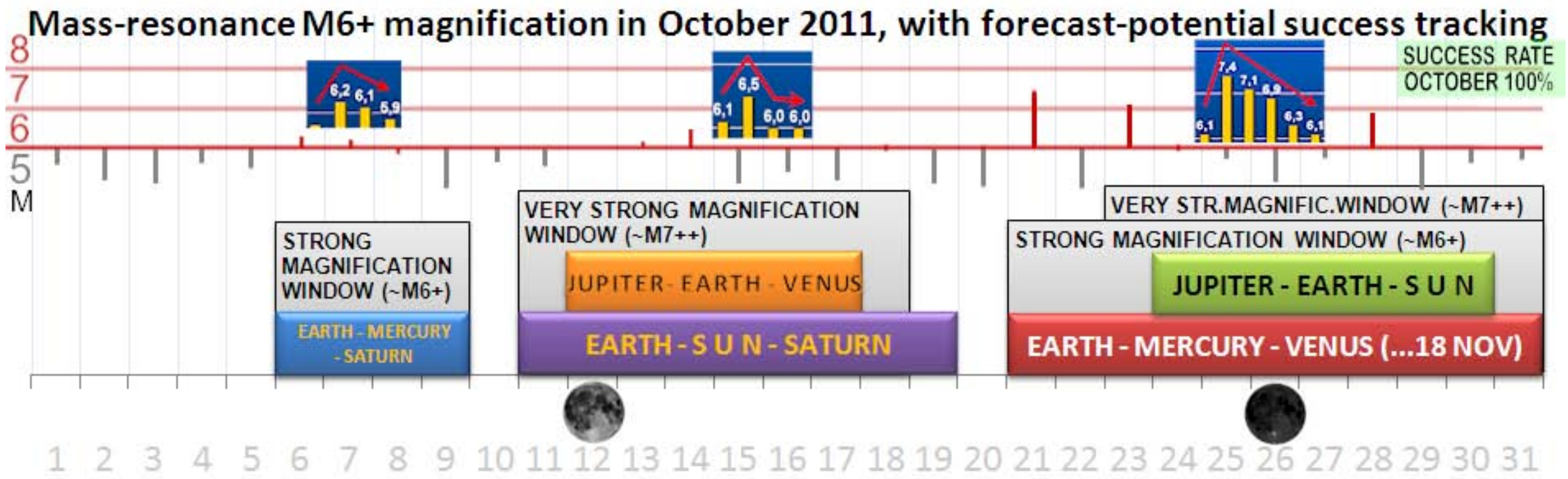

Figure 14: October, 2011 hyperresonance forecast of Earth's M6+ magnified shaking. Daily-strongest M5+ earthquakes tick-marked. Full Moon interposing M5+ seismicity resulted in interference (attenuation) of the resonance M7+ magnification during a multiple alignment involving the Sun. Importantly, all M6+ patterns as predicted. Not only did the first 5 days of October without long alignments go calm in the sense of M6+ seismicity, but the only day in October without M5+ earthquakes at all was the day of the Full Moon. That is a 1-in-31 chance, or mere 3%, for such a lunar effect to be a coincidence and not due to physical principle of the georesonator. On top of this revealing statistics, and again in accordance with the georesonator concept, the only interval in October with a multiple long alignment not interfered by a fourth body, has indeed seen M7+ quakes. (This was also the only M7+ magnification episode of the entire month.) Furthermore, all three M6+ seismicity episodes (as distinguished by always the same pattern of earthquake strength's increase-peaking-decrease) in October corresponded one-on-one to only three (independent) intervals with long alignments of that month.

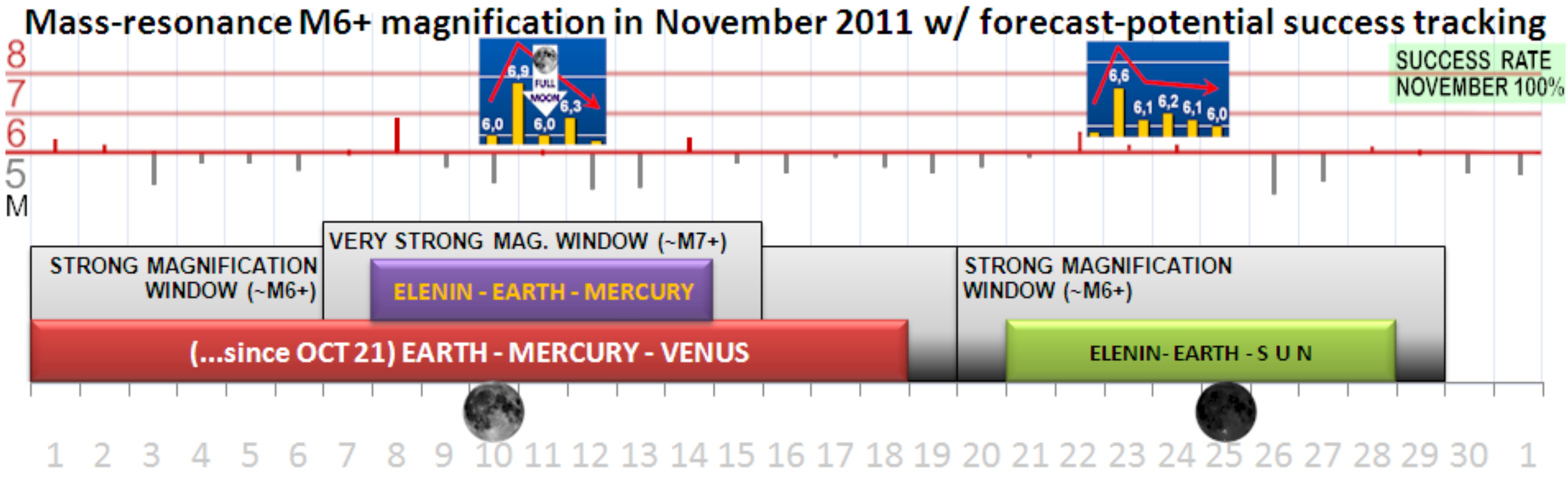

Figure 15: November, 2011, hyperresonance forecast of Earth's M6+ magnified shaking. Daily-strongest M5+ earthquakes tick-marked. New Moon interposing M5+ seismicity resulted in interference (attenuation) of the resonance M6+ magnification during an alignment involving the Sun. Importantly, all M6+ patterns as predicted. Not only did the first days of November saw M6+ seismicity that continued the last M6+ episode's pattern from October, but the only day in November without M5+ earthquakes at all was the day of the New Moon. That is a 1-in-31 chance, or mere 3%, for such a lunar effect to be a coincidence and not due to physical principle of the georesonator. On top of this revealing statistics, and again in accordance with the georesonator concept, the only interval in November with a multiple long alignment not interfered by a fourth body, has indeed seen ~M7+ seismicity. (This was also the only ~M7+ magnification episode of the entire month.) Furthermore, both M6+ seismicity episodes (as distinguished by always the same pattern of earthquake strength's increase-peaking-decrease) in November did correspond one-on-one to the only two (independent) intervals with long alignments of that month. Due to Full Moon interposing, the 11 November, M6.0, earthquake was significantly weaker than expected (i.e., than the usual pattern); same was observed in September, Figure 16.

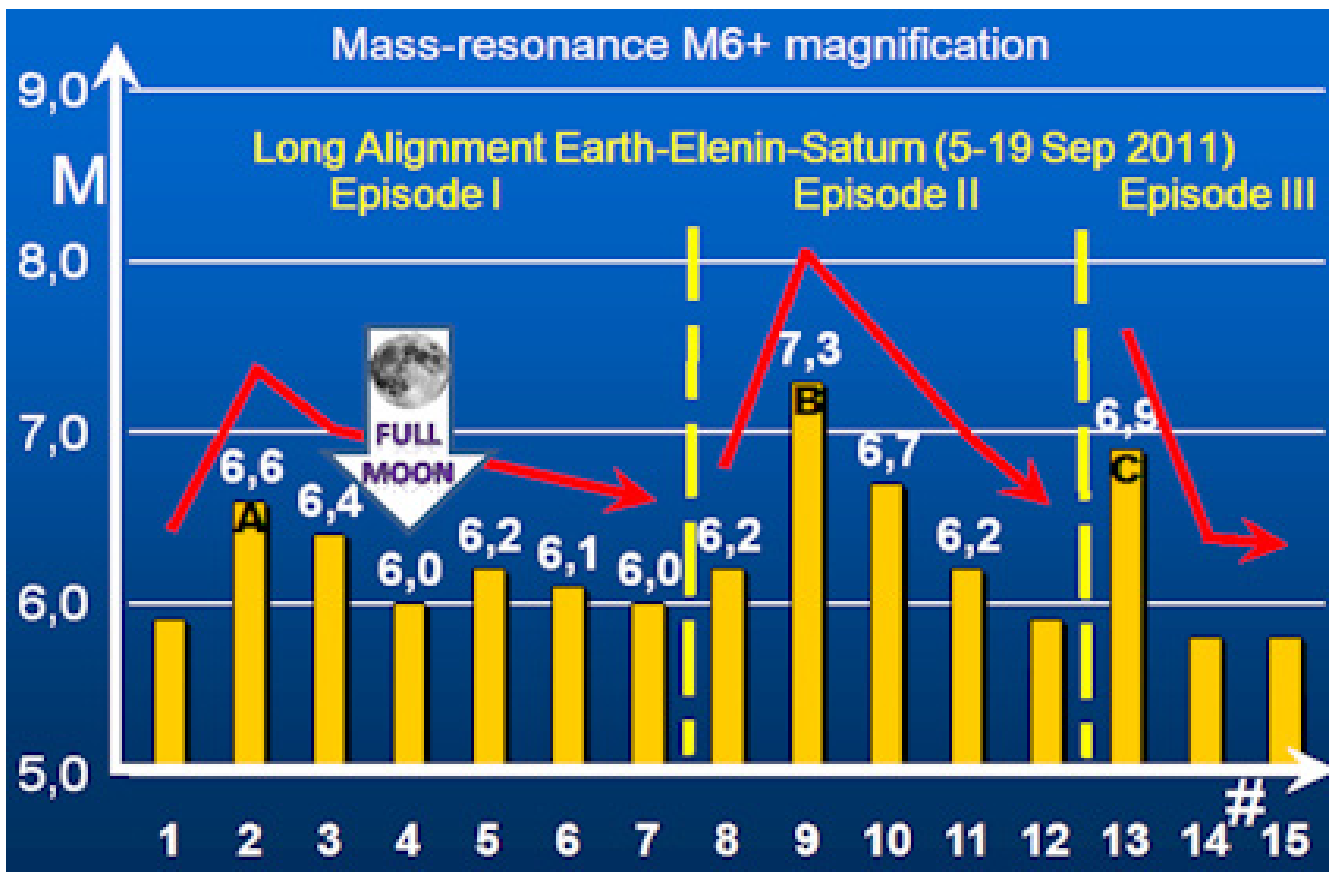

Figure 16. Magnitudes of consecutive strong earthquakes during a very long alignment in September 2011.

## A: Discarded data in April-November analyses

| DATE | Mw | REASON FOR EXCLUSION |
|---|---|---|
| May 21 | 6.1 | 3+ DAYS W/O ALIGNMENT; OTHERS REPORTED AS M6- (5.7; 5.8)** |
| July 19 | 6.1 | 3+ DAYS W/O ALIGNMENT (6.0; 6.1) |
| August 24 | 6.2 | OTHERS REPORTED AS M6- AFTERSHOCK (6.0; 6.2) |
| September 15 | 6.0 | OTHERS REPORTED AS M6- AFTERSHOCK (5.4; 5.4) |
| October 27 | 6.0 | OTHERS REPORTED AS M6- (5.9; 6.0); UNRELIABLE DEPTH (>600km) |

* Note that, under the georesonance concept criteria, only M(6.0+5%)– outliers can be excluded from considerations.
** Magnitudes in parentheses are the estimates by GFZ and EMSC, respectively.

## B: Final data in October-November forecast testing (sources' updates as of 30 November)

| Date | USGS | GFZ | EMSC | Mw **Final** | Comments |
|---|---|---|---|---|---|
| 6 Oct | 5.9 | 5.9 | 6.2 | **6.2** | Cannot discard* |
| 7 Oct | 6.1 | 6.0 | 6.1 | **6.1** | |
| 13 Oct | 6.1 | 6.1 | 6.1 | **6.1** | |
| 14 Oct | 6.5 | 6.3 | 6.5 | **6.5** | |
| 14 Oct | 6.0 | 5.9 | 6.0 | **6.0** | |
| 18 Oct | 6.1 | 5.9 | 6.0 | **6.0** | |
| 21 Oct | 6.1 | 6.1 | 6.1 | **6.1** | |
| 21 Oct | 7.4 | 7.3 | 7.4 | **7.4** | |
| 23 Oct | 7.1 | 7.1 | 7.2 | **7.1** | |
| 28 Oct | 6.9 | 6.9 | 6.9 | **6.9** | |
| 1 Nov | 6.3 | 6.2 | 6.3 | **6.3** | |
| 2 Nov | 6.1 | 6.1 | 6.3 | **6.1** | |
| 7 Nov | 6.0 | 6.1 | 6.0 | **6.0** | |
| 8 Nov | 6.9 | 6.9 | 6.9 | **6.9** | |
| 11 Nov | 6.0 | 5.9 | 6.0 | **6.0** | |
| 14 Nov | 6.3 | 6.3 | 6.3 | **6.3** | |
| 22 Nov | 6.6 | 6.5 | 6.6 | **6.6** | |
| 23 Nov | 6.1 | 6.1 | 6.1 | **6.1** | |
| 24 Nov | 6.2 | 6.2 | 6.2 | **6.2** | |
| 28 Nov | 6.4 | 6.1 | 6.1 | **6.1** | |

Table 7. **Section A** - List of all M6+ earthquakes (USGS) excluded* from the April-November analyses, and reason. Note that only five earthquakes were omitted from the analyses over those seven months, or less than one per month on average. **Section B** – Final magnitudes of M6+ earthquakes in the October-November real-time forecast testing, Figures 14 and 15 respectively, and how those magnitudes were determined: by comparing the respective USGS, GFZ, and EMSC estimates to discard outliers. The GFZ data from: http://geofon.gfz-potsdam.de/eqinfo/list.php, the EMSC data from: http://www.emsc-csem.org/Earthquake/.

# APPENDIX E

**Demonstration of 100% accurate prediction of Earth's M6+ magnified oscillation due to long astronomical alignments, based on georesonator concept from 6+ months of successive real-time tracking of the prediction from Oct 2011–Jun 2012** (www.seismo.info).

The 6-months-prediction of resonance magnification and pattern detection ended 11 April 2012 (v.6) in a spectacular fashion: with the first-ever recorded singlet of two M8+ earthquakes (at same time *and* locale; here within less than 2 hrs and less than 2° of latitude and longitude from each other), Fig.18. The singlet was not due to any stress accumulation, for such a short distance *and* time span would mean a M11+ earthquake had occurred (for example, a ~M8+ doublet of 12 September 2007 in similar settings as this singlet, while less energetic, took half a day, Fig.18). Since such stress accumulation is unattainable on the Earth, such M8+ singlet, which occur on average less than once a century, can only arise due to the Earth's simultaneous very long alignments. <u>This author hereby states that he purposely and tacitly selected publication date of this preprint's version 1 so as to make it precede the spectacular singlet for one year. As based on the 2007 doublet, this anniversary serves as a proof-of-point for georesonator, while highlighting the discoverer's priority claim.</u> Then v.6 was the necessary one.

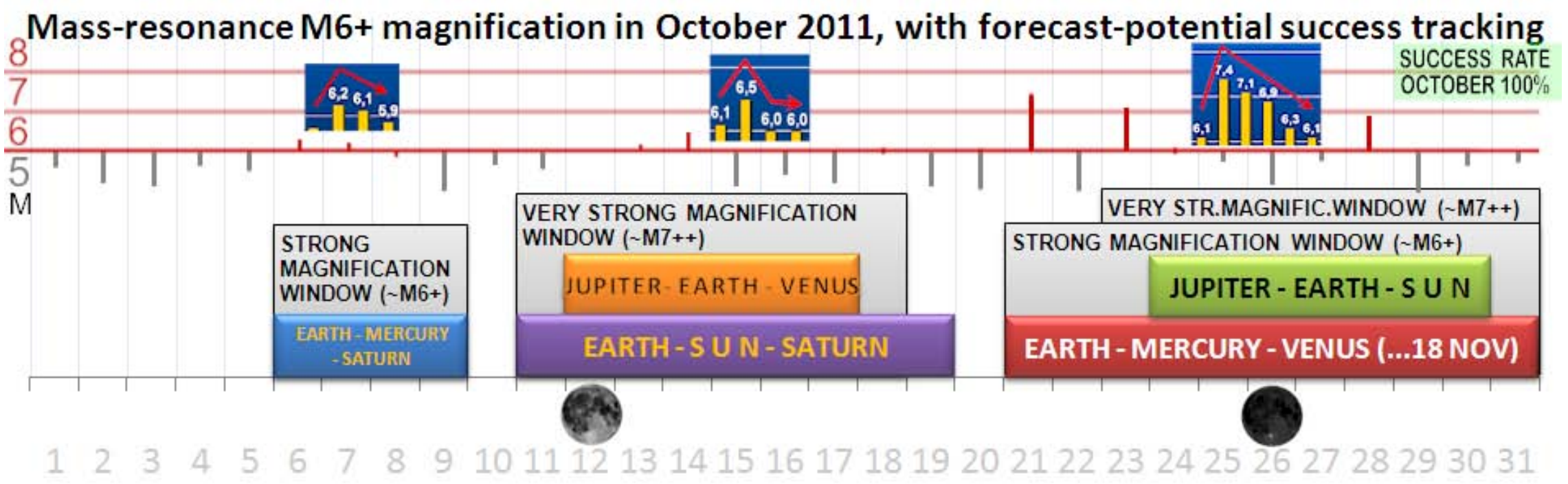

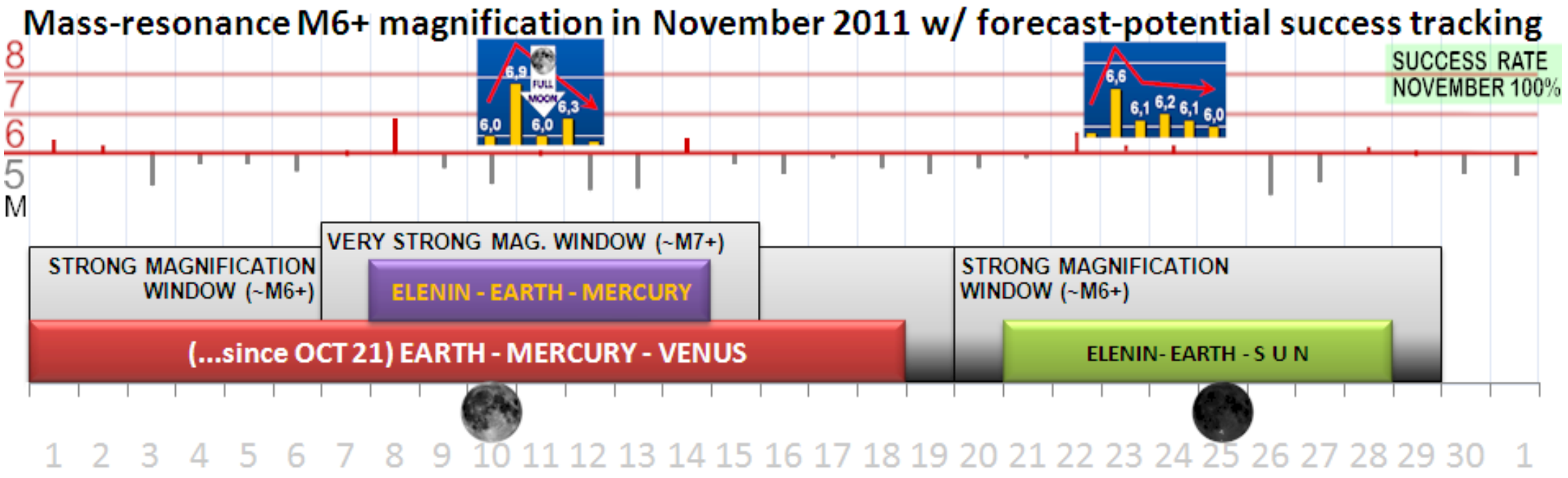

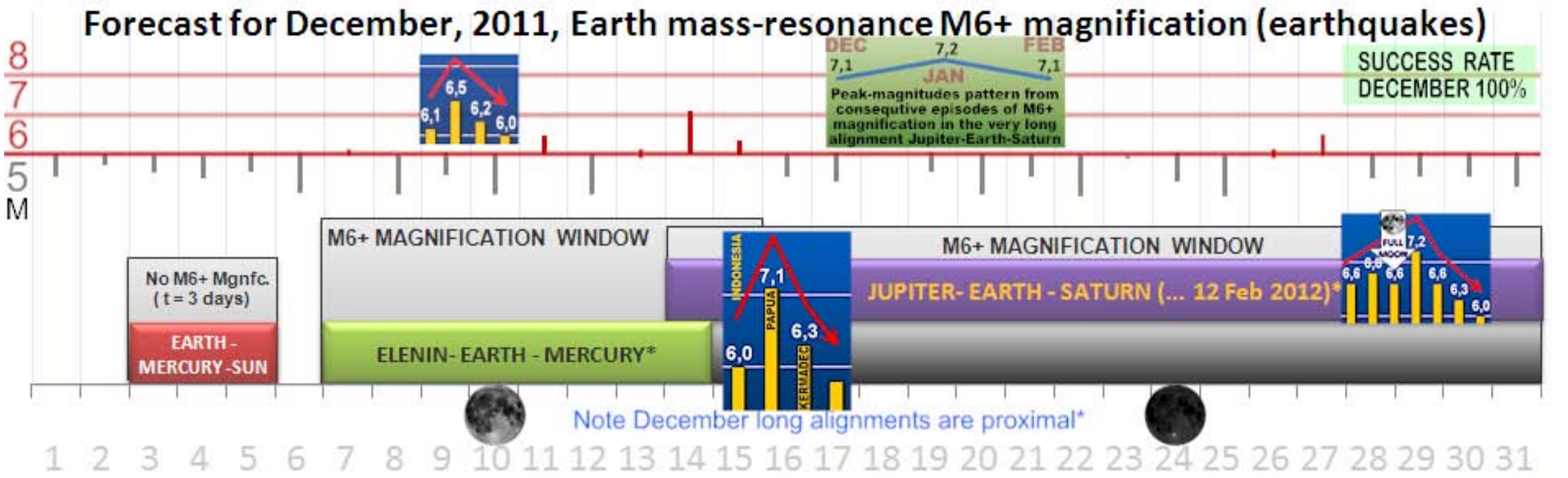

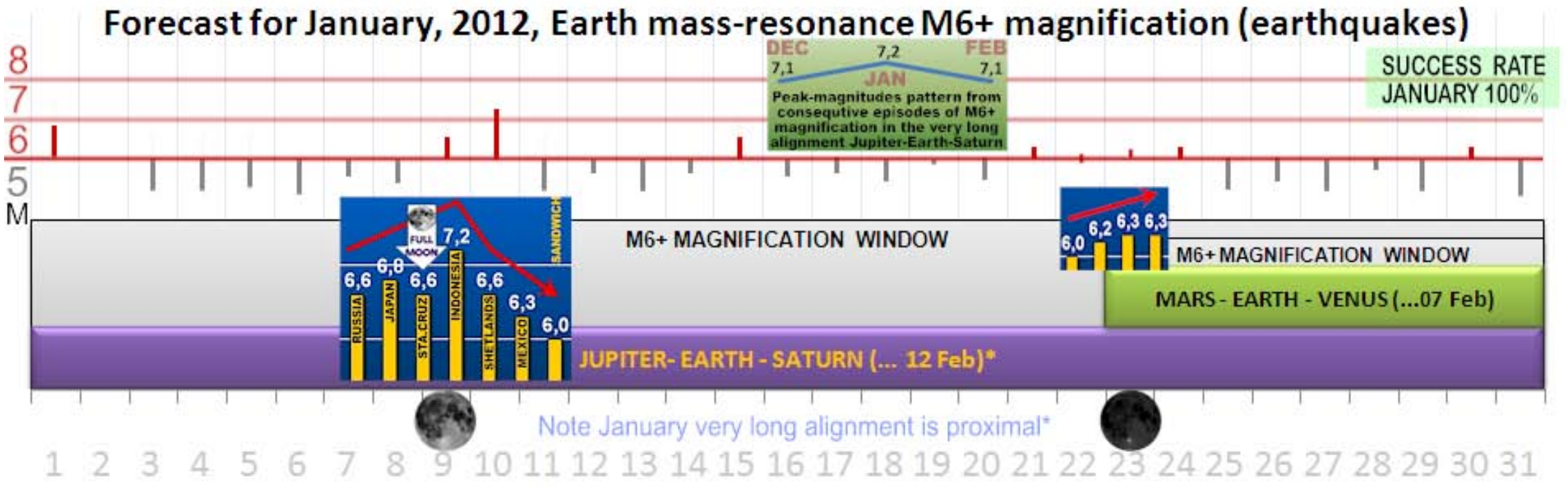

Forecast for January, 2012, Earth mass-resonance M6+ magnification (earthquakes)
SUCCESS RATE JANUARY 100%
Peak-magnitudes pattern from consequtive episodes of M6+ magnification in the very long alignment Jupiter-Earth-Saturn
M6+ MAGNIFICATION WINDOW
M6+ MAGNIFICATION WINDOW
MARS - EARTH - VENUS (...07 Feb)
JUPITER- EARTH - SATURN (... 12 Feb)*
Note January very long alignment is proximal*

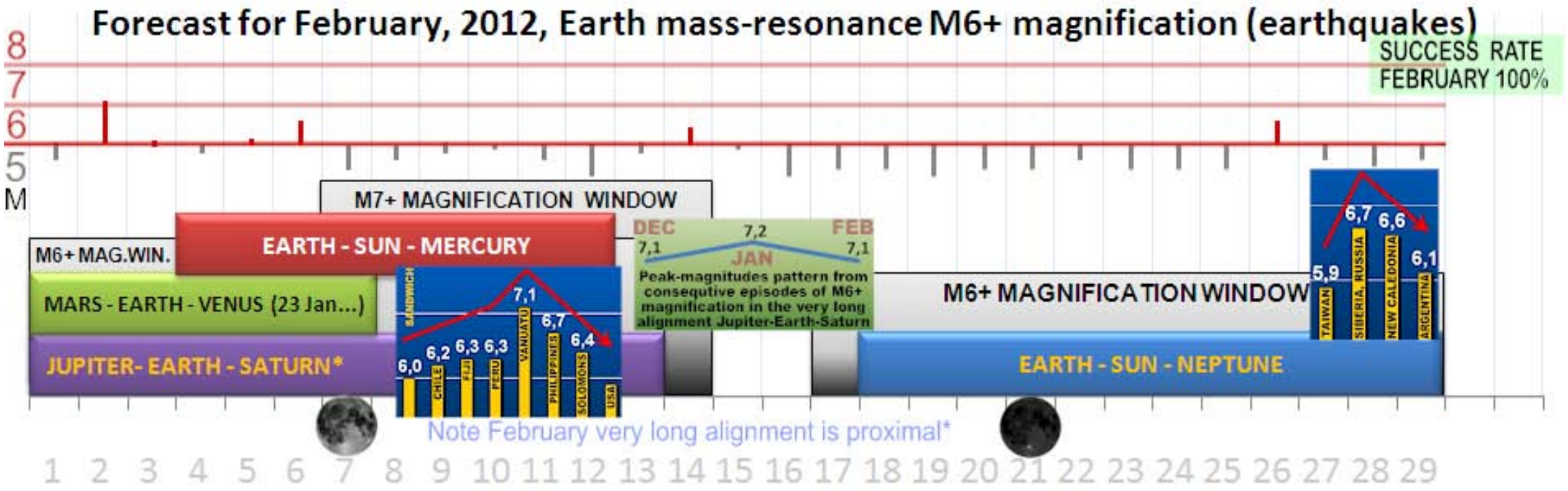

Forecast for February, 2012, Earth mass-resonance M6+ magnification (earthquakes)
SUCCESS RATE FEBRUARY 100%
M7+ MAGNIFICATION WINDOW
M6+ MAG.WIN.
EARTH - SUN - MERCURY
MARS - EARTH - VENUS (23 Jan...)
JUPITER- EARTH - SATURN*
M6+ MAGNIFICATION WINDOW
EARTH - SUN - NEPTUNE
Note February very long alignment is proximal*

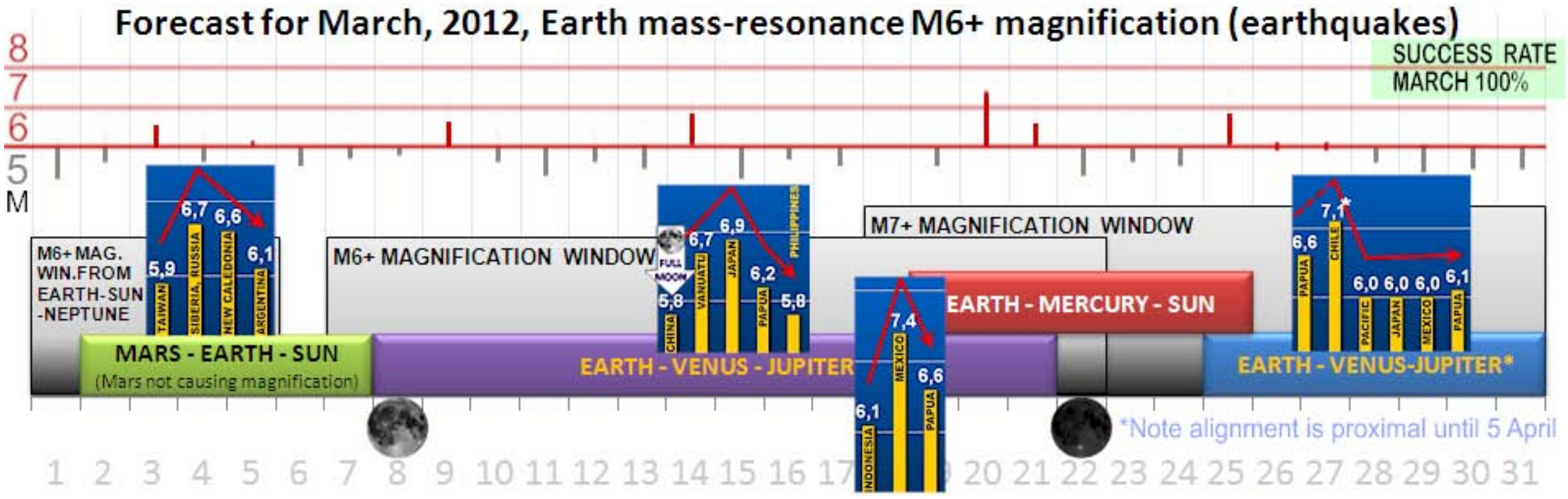

Forecast for March, 2012, Earth mass-resonance M6+ magnification (earthquakes)
SUCCESS RATE MARCH 100%
M6+ MAG. WIN.FROM EARTH-SUN -NEPTUNE
M6+ MAGNIFICATION WINDOW
M7+ MAGNIFICATION WINDOW
EARTH - MERCURY - SUN
MARS - EARTH - SUN
(Mars not causing magnification)
EARTH - VENUS - JUPITER
EARTH - VENUS-JUPITER*
*Note alignment is proximal until 5 April

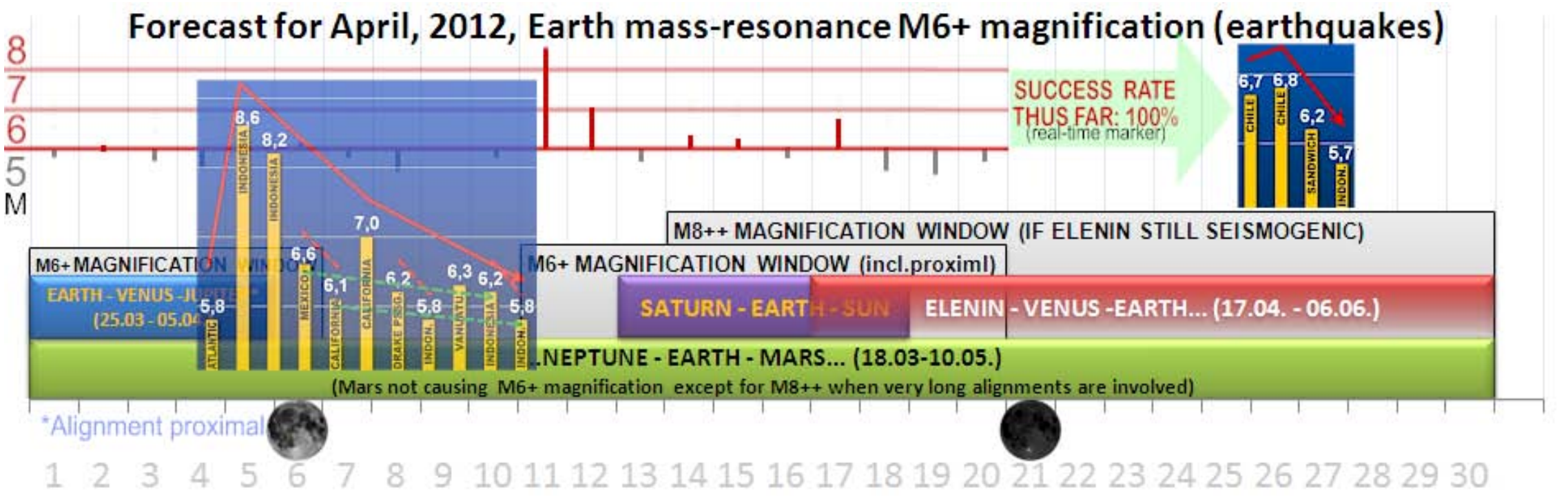

Forecast for April, 2012, Earth mass-resonance M6+ magnification (earthquakes)
SUCCESS RATE THUS FAR: 100%
(real-time marker)
M8++ MAGNIFICATION WINDOW (IF ELENIN STILL SEISMOGENIC)
M6+ MAGNIFICATION WINDOW (incl.proximl)
SATURN - EARTH - SUN
ELENIN - VENUS -EARTH... (17.04. - 06.06.)
NEPTUNE - EARTH - MARS... (18.03-10.05.)
(Mars not causing M6+ magnification except for M8++ when very long alignments are involved)
*Alignment proximal

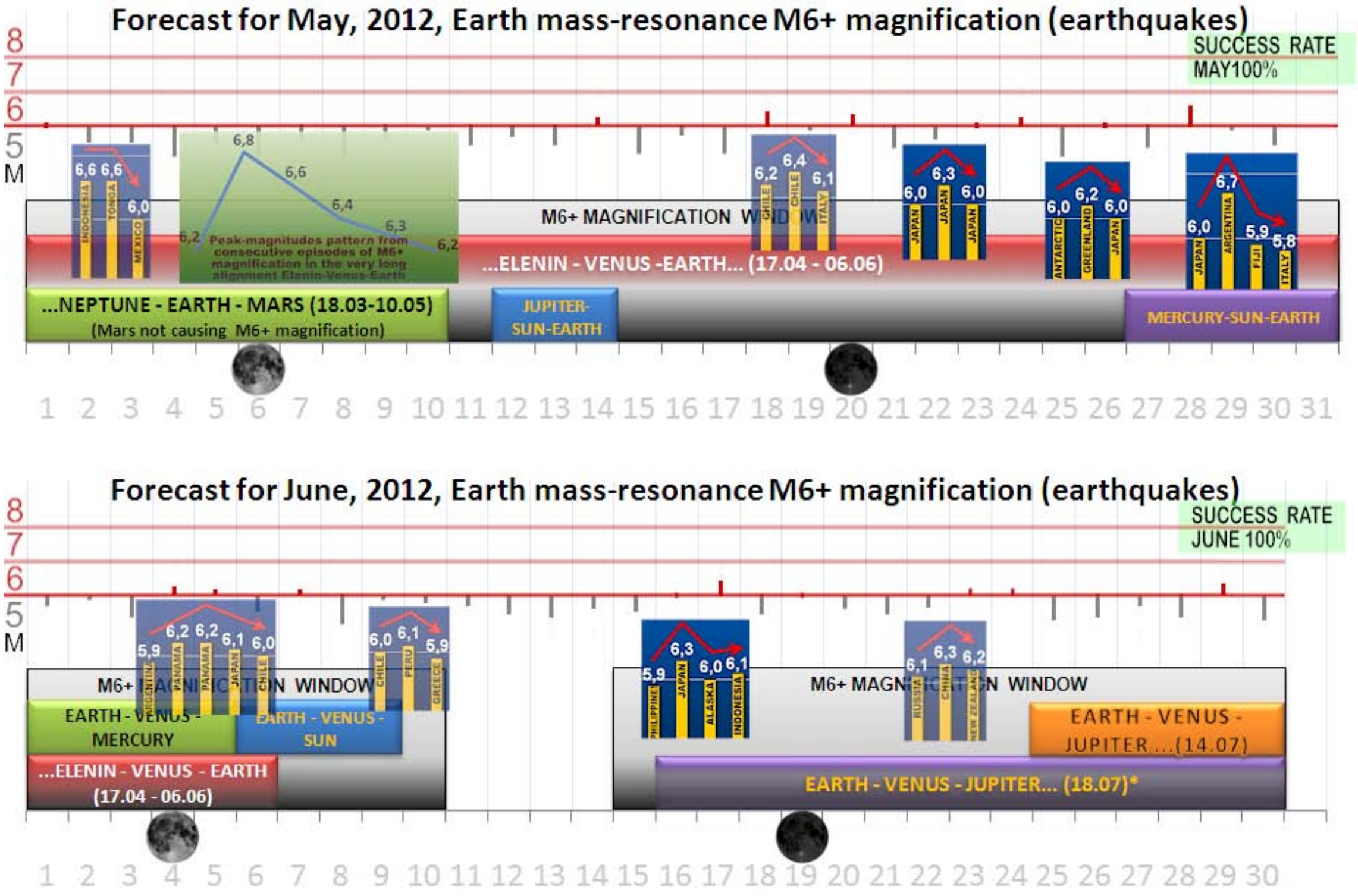

Figure 17: The October 2011–June 2012, 100%-correct demonstration of the increase-peaking-decrease pattern in the Earth's magnified oscillation as due to Earth's long astronomical alignments. To eliminate outliers, the earthquake magnitudes were selected amongst three (USGS, GFZ, EMSC) estimates. For proper legends under each of the above panels, consult the legends under Figures 14 and 15. Note that *proximal long alignments*, marked with an *, occur when heavenly bodies are non-strictly aligned; that is, when bodies never align themselves or they do on occasion, all while mostly staying aligned to within the ±5° window.

## A: Discarded data in April 2011–June 2012 analyses

| DATE | M | REASON FOR EXCLUSION |
|---|---|---|
| 1) May 21 | 6.1 | 3+ DAYS W/O ALIGNMENT; OTHERS REPORTED AS M6- (5.7; 5.8)** |
| 2) July 19 | 6.1 | 3+ DAYS W/O ALIGNMENT (6.0; 6.1) |
| 3) August 24 | 6.2 | OTHERS REPORTED AS M6- AFTERSHOCK (6.0; 6.2) |
| 4) September 15 | 6.0 | OTHERS REPORTED AS M6- AFTERSHOCK (5.4; 5.4) |
| 5) October 27 | 6.0 | OTHERS REPORTED AS M6- (5.9; 6.0); UNRELIABLE DEPTH (>600km) |
| 6) Feb 3, 2012 | 6.0 | AFTERSHOCK OF M7.1 VANUATU OF 02 FEB, proximate alignment |
| 6) Feb 5, 2012 | 6.1 | AFTERSHOCK OF M7.1 VANUATU OF 02 FEB 2012; proximate alignment, (5.6; 5.7)*** |
| 6) Feb 5, 2012 | 6.0 | AFTERSHOCK OF M7.1 VANUATU OF 02 FEB 2012, proximate alignment, (6.2; 6.1)*** |
| 7) Feb 6, 2012 | 6.0 | AFTERSHOCK OF M6.7 PHILIPPINES QUAKE OF 06 FEB 2012, proximate alignment |
| 8) Mar 14, 2012 | 6.1 | AFTERSHOCK OF M6.9 JAPAN QUAKE OF 14 MAR 2012 (6.0; 6.0)*** |
| 9) April 11, 2012 | 6.0 | OTHERS TOO REPORT AS M6.1-, REMAINING REPORT AN OUTLIER (5.9; 6.3) |

* Note that, under the georesonance concept criteria, only M(6.0+5%)– outliers can be excluded from considerations.
** Magnitudes in parentheses are the estimates by GFZ and EMSC, respectively.

**B: Final data in October 2011–April 2012 forecast of ~M6+ magnification of Earth's resonance**

| Date | USGS | GFZ | EMSC | Final | Comments |
|---|---|---|---|---|---|
| 6 Oct | 5.9 | 5.9 | 6.2 | **6.2** | (cannot discard*) Argentina |
| 7 Oct | 6.1 | 6.0 | 6.1 | **6.1** | Kermadec Isles |
| 13 Oct | 6.1 | 6.1 | 6.1 | **6.1** | Indonesia |
| 14 Oct | 6.5 | 6.3 | 6.5 | **6.5** | Papua New Guinea |
| 14 Oct | 6.0 | 5.9 | 6.0 | **6.0** | Russia |
| 18 Oct | 6.1 | 5.9 | 6.0 | **6.0** | Papua New Guinea |
| 21 Oct | 6.1 | 6.1 | 6.1 | **6.1** | Japan |
| 21 Oct | 7.4 | 7.3 | 7.4 | **7.4** | Kermadec Isles |
| 23 Oct | 7.1 | 7.1 | 7.2 | **7.1** | Turkey |
| 28 Oct | 6.9 | 6.9 | 6.9 | **6.9** | Peru |
| 1 Nov | 6.3 | 6.2 | 6.3 | **6.3** | Revilla Gigedo Isles, Mexico |
| 2 Nov | 6.1 | 6.1 | 6.3 | **6.1** | Pacific-Antarctic Ridge |
| 7 Nov | 6.0 | 6.1 | 6.0 | **6.0** | Nicaragua |
| 8 Nov | 6.9 | 6.9 | 6.9 | **6.9** | Taiwan |
| 11 Nov | 6.0 | 5.9 | 6.0 | **6.0** | Southern East Pacific Rise |
| 14 Nov | 6.3 | 6.3 | 6.3 | **6.3** | Molucca Sea, Indonesia |
| 22 Nov | 6.6 | 6.5 | 6.6 | **6.6** | Bolivia |
| 23 Nov | 6.1 | 6.1 | 6.1 | **6.1** | Japan |
| 24 Nov | 6.2 | 6.2 | 6.2 | **6.2** | Japan |
| 28 Nov | 6.4 | 6.1 | 6.1 | **6.1** | Papua New Guinea |
| 29 Nov | 6.0 | 6.0 | 6.0 | **6.0** | Ascension Island, UK territory |
| 07 Dec | 6.1 | 6.0 | 6.1 | **6.1** | Chile |
| 11 Dec | 6.5 | 6.5 | 6.7 | **6.5** | Mexico |
| 11 Dec | 6.2 | 6.2 | 6.3 | **6.2** | Sandwich Isles |
| 13 Dec | 6.0 | 6.0 | 6.1 | **6.0** | Indonesia |
| 14 Dec | 7.1 | 7.1 | 7.1 | **7.1** | Papua New Guinea |
| 15 Dec | 6.3 | 5.9 | 6.0 | **6.3** | (cannot discard*) Kermadec Isles |
| 26 Dec | 6.0 | 6.0 | 6.1 | **6.0** | Tonga |
| 27 Dec | 6.6 | 6.6 | 6.5 | **6.6** | Russia |
| 1 Jan | 6.8 | 6.7 | 7.0 | **6.8** | Japan |
| 9 Jan | 6.6 | 6.4 | 6.6 | **6.6** | Sta. Cruz |
| 10 Jan | 7.2 | 7.1 | 7.2 | **7.2** | Indonesia |
| 15 Jan | 6.6 | 6.5 | 6.7 | **6.6** | Shetland Isles |
| 21 Jan | 6.3 | 6.2 | 6.3 | **6.3** | Mexico |
| 22 Jan | 6.0 | 6.0 | 6.0 | **6.0** | Sandwich Isles |
| 23 Jan | 6.2 | 6.1 | 6.2 | **6.2** | Chile |
| 24 Jan | 6.3 | 6.3 | 6.4 | **6.3** | Fiji |
| 30 Jan | 6.3 | 6.4 | 6.3 | **6.3** | Peru |
| 2 Feb | 7.1 | 7.0 | 7.1 | **7.1** | Vanuatu |
| 6 Feb | 6.7 | 6.7 | 6.7 | **6.7** | Philippines |
| 14 Feb | 6.4 | 6.3 | 6.4 | **6.4** | Solomon Isles |
| 15 Feb | 6.0 | 5.7 | 5.9 | **5.9** | Oregon, USA |
| 26 Feb | 5.9 | 5.8 | 5.9 | **5.9** | Taiwan |
| 26 Feb | 6.7 | 6.5 | 6.8 | **6.7** | Siberia, Russia |
| 3 Mar | 6.6 | 6.6 | 6.6 | **6.6** | New Caledonia |
| 5 Mar | 6.1 | 6.1 | 6.1 | **6.1** | Argentina |
| 8 Mar | 5.8 | 5.8 | 5.8 | **5.8** | Xinjiang, China |
| 9 Mar | 6.7 | 6.7 | 6.7 | **6.7** | Vanuatu |
| 14 Mar | 6.9 | 6.9 | 6.9 | **6.9** | Japan |

| Date | USGS | GFZ | EMSC | Final | Comments |
|---|---|---|---|---|---|
| 14 Mar | 6.2 | 6.2 | 6.2 | **6.2** | Papua New Guinea |
| 16 Mar | 5.7 | 5.8 | 5.8 | **5.8** | Philippines |
| 20 Mar | 6.2 | 6.1 | 6.1 | **6.1** | Indonesia |
| 20 Mar | 7.4 | 7.3 | 7.4 | **7.4** | Mexico |
| 21 Mar | 6.6 | 6.6 | 6.7 | **6.6** | Papua New Guinea |
| 25 Mar | 7.1 | 7.1 | 7.1 | **7.1** | Chile |
| 26 Mar | 6.0 | 6.0 | 6.0 | **6.0** | Pacific Rise |
| 27 Mar | 6.0 | 6.0 | 6.0 | **6.0** | Japan |
| 2 Apr | 6.0 | 6.0 | 6.0 | **6.0** | Mexico |
| 6 Apr | 6.2 | 6.1 | 6.1 | **6.1** | Papua New Guinea |
| 11 Apr | 8.6 | 8.6 | 8.6 | **8.6** | Indonesia |
| 11 Apr | 8.2 | 8.2 | 8.0 | **8.2** | Indonesia |
| 11 Apr | 6.5 | 6.6 | 6.7 | **6.6** | Mexico |
| 12 Apr | 6.2 | 6.1 | 6.0 | **6.1** | Gulf of California |
| 12 Apr | 6.9 | 7.0 | 7.0 | **7.0** | Gulf of California |
| 14 Apr | 6.2 | 6.2 | 6.2 | **6.2** | Drake Passage |
| 14 Apr | 5.9 | 5.8 | 5.8 | **5.8** | Indonesia |
| 14 Apr | 6.5 | 6.3 | 6.3 | **6.3** | Vanuatu |
| 15 Apr | 6.2 | 6.2 | 6.2 | **6.2** | Indonesia |
| 16 Apr | 5.8 | 5.8 | 5.7 | **5.8** | Indonesia |
| 17 Apr | 6.7 | 6.7 | 6.7 | **6.7** | Chile |
| 17 Apr | 6.8 | 6.8 | 7.0 | **6.8** | Chile |
| 17 Apr | 6.2 | 6.2 | 6.2 | **6.2** | Sandwich Isles |
| 21 Apr | 6.7 | 6.5 | 6.6 | **6.6** | Indonesia |
| 28 Apr | 6.6 | 6.6 | 6.6 | **6.6** | Tonga |
| 1 May | 6.0 | 5.9 | 6.0 | **6.0** | Mexico |
| 14 May | 6.2 | 6.2 | 6.2 | **6.2** | Chile |
| 18 May | 6.2 | 6.4 | 6.4 | **6.4** | Chile |
| 20 May | 6.0 | 6.1 | 6.1 | **6.1** | Italy |
| 20 May | 5.9 | 6.2 | 6.0 | **6.0** | Japan |
| 20 May | 6.4 | 6.3 | 6.3 | **6.3** | Japan |
| 23 May | 6.0 | 5.9 | 6.0 | **6.0** | Japan |
| 23 May | 5.9 | 6.0 | 6.0 | **6.0** | West Indian-Antarctic Ridge |
| 24 May | 6.2 | 6.2 | 6.2 | **6.2** | Greenland Sea |
| 26 May | 6.0 | 5.5 | 6.0 | **6.0** | Japan |
| 28 May | 6.7 | 6.6 | 6.7 | **6.7** | Argentina |
| 2 Jun | 5.9 | 5.9 | 6.0 | **5.9** | Argentina |
| 4 Jun | 6.3 | 6.2 | 6.2 | **6.2** | Panama |
| 4 Jun | 6.3 | 6.2 | 6.2 | **6.2** | Panama |
| 5 Jun | 6.1 | 6.0 | 6.1 | **6.1** | Japan |
| 7 Jun | 6.0 | 6.0 | 5.9 | **6.0** | Chile |
| 7 Jun | 6.1 | 6.1 | 6.1 | **6.1** | Peru |
| 10 Jun | 6.0 | 5.9 | 5.9 | **5.9** | Greece |
| 16 Jun | 5.9 | 5.9 | 5.8 | **5.9** | Philippines |
| 17 Jun | 6.3 | 6.3 | 6.4 | **6.3** | Japan |
| 19 Jun | 6.0 | 5.9 | 6.0 | **6.0** | Aleutian Isles/Alaska |
| 23 Jun | 6.1 | 6.1 | 6.1 | **6.1** | Indonesia |
| 24 Jun | 6.0 | 6.1 | 6.1 | **6.1** | Kamchatka/Russia |
| 29 Jun | 6.2 | 6.3 | 6.3 | **6.3** | China |

Table 8. Same as Table 7 but with data through June, 2012.

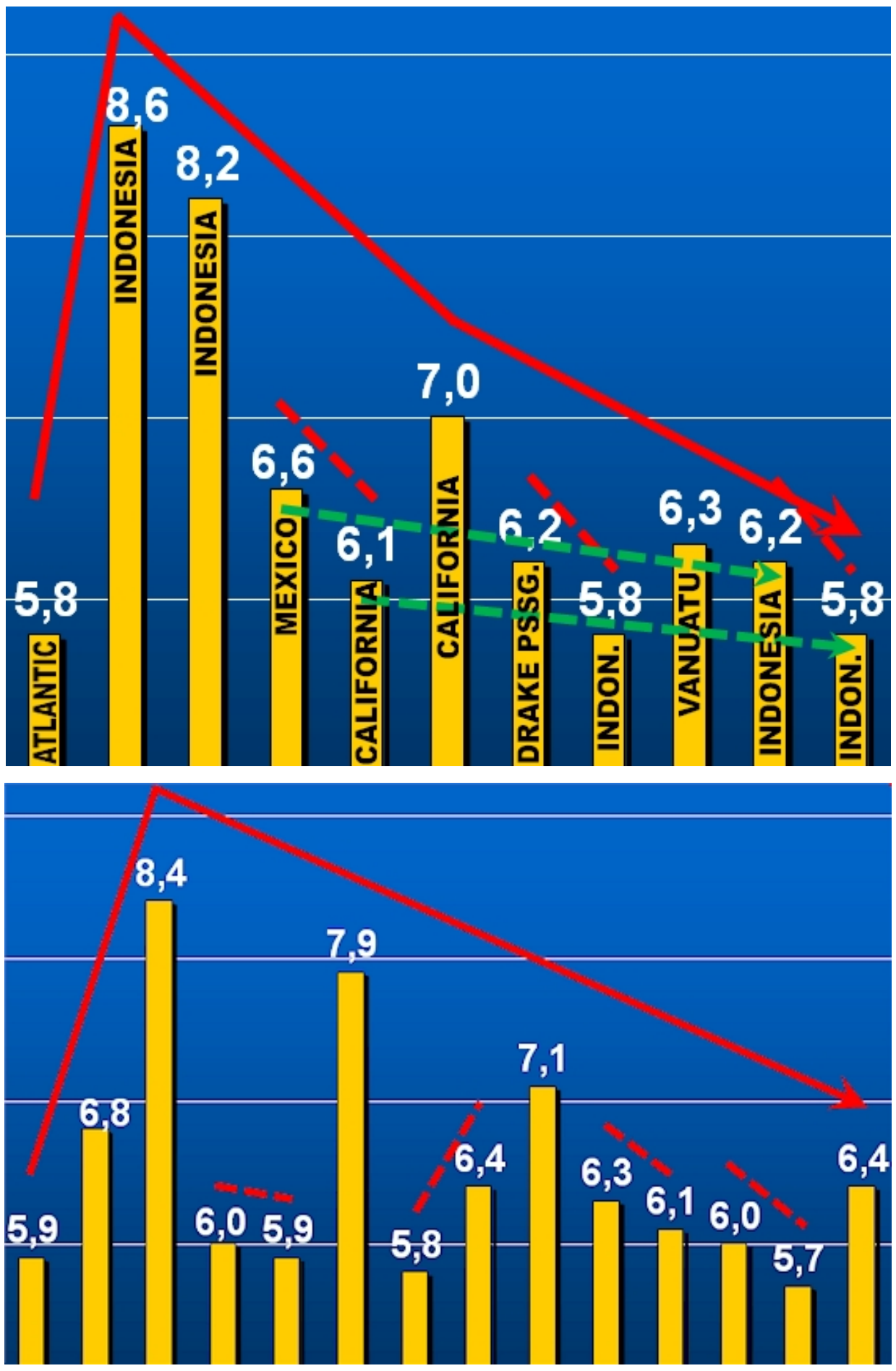

Figure 18: The 11 April 2012, spectacular M8+ singlet (top panel; Table 9), shown on the last panel of Fig.17 as due to very long alignments with the Mars (and the Neptune) and, simultaneously, with the Saturn and the Sun. The above singlet proves the georesonator conclusively: while the singlet's former (M8.6) earthquake produced the main tone, the latter (M8.2) resulted in resonant overtones (dashed lines), as expected in a georesonator but not in simplistic stress/strain models. Also, the fact that overtones can be of an in-step type (i.e., magnitude change is regular), corroborates the georesonator conclusively: the overtones' step is always random under stress/strain models, as such models are determined by the Earth's complex interior. The above shown singlet on the other hand is completely regular, unlike say the 12 September 2007 doublet (bottom panel; Table 9) due to very long alignments with the Mars (and the Jupiter) and, simultaneously, with the Venus and the Elenin; also in Indonesia but which took half a day to complete thereby losing its character of a singlet (no regularities can be seen in the overtone-changes of magnitudes, bottom panel). A much simpler, i.e., extraterrestrial as the only alternative, mechanism is responsible for seismogenesis as well as tectonogenesis then. This conclusion is also supported by a great location diversity of the events that obey the above main tone and overtones as due to a singlet; so the mechanism is also global in character. Note Indonesia seems very vulnerable to rare (rarer than per decade) very long alignments combined with those to the Mars.

| date | time | location | | d | M | M | M | M |
|---|---|---|---|---|---|---|---|---|
| 2007 09 07 | 044645.16 | -56.08 | -123.31 | 10 | 5.9 | | 5.9 | 5.9 |
| 2007 09 10 | 014911.78 | 2.97 | -77.96 | 15 | 6.8 | ~~7.0~~ | 6.8 | 6.8 |
| 2007 09 12 | 111026.83 | -4.44 | 101.37 | 34 | 8.5 | ~~8.0~~ | 8.4 | 8.4 |
| 2007 09 12 | 144005.73 | -3.16 | 101.46 | 35 | 5.9 | ~~6.4~~ | 6.0 | 6.0 |
| 2007 09 12 | 163703.92 | -3.14 | 101.40 | 35 | 5.8 | ~~6.1~~ | 5.9 | 5.9 |
| 2007 09 12 | 234903.72 | -2.62 | 100.84 | 35 | 7.9 | 8.0 | 7.8 | 7.9 |
| 2007 09 13 | 012634.42 | -1.90 | 99.82 | 16 | 5.7 | ~~6.4~~ | 5.8 | 5.8 |
| 2007 09 13 | 023003.30 | -1.69 | 99.67 | 28 | 6.5 | 6.3 | ~~6.0~~ | 6.4 |
| 2007 09 13 | 033528.72 | -2.13 | 99.63 | 22 | 7.0 | 7.2 | 7.1 | 7.1 |
| 2007 09 13 | 094845.13 | 3.80 | 126.34 | 26 | 6.3 | ~~6.6~~ | 6.3 | 6.3 |
| 2007 09 13 | 160916.87 | -3.17 | 101.52 | 53 | 6.0 | ~~6.5~~ | 6.2 | 6.1 |
| 2007 09 13 | 200638.92 | -57.62 | -147.30 | 10 | 6.0 | | 5.9 | 6.0 |
| 2007 09 14 | 054645.70 | -15.46 | -176.18 | 35 | 5.7 | | 5.7 | 5.7 |
| 2007 09 14 | 060132.27 | -4.07 | 101.17 | 23 | 6.4 | ~~6.6~~ | 6.4 | 6.4 |
| Table 9. USGS data, for bottom panel of Fig.18. | | | | | | GFZ | EMSC | final |